\documentclass[rspublic, superscriptaddress, showpacs, notitlepage, twocolumn]{revtex4-1}

\usepackage{amsmath, amsfonts}
\usepackage{xspace}

\usepackage{caption}
\usepackage{float}
\usepackage{amssymb}
\usepackage{color}
\usepackage[usenames,dvipsnames]{xcolor}

 \usepackage{algpseudocode} 
\usepackage{hyperref}
\usepackage{amsthm}
\usepackage{graphicx}
\usepackage{dcolumn}
\usepackage{bm}
\usepackage{subfigure}
\usepackage{amssymb}
\usepackage{verbatim}
\usepackage{amscd}
\usepackage{amssymb}

\usepackage{setspace}
\usepackage[]{graphicx}        
\usepackage{amsthm}
\usepackage{natbib}
\usepackage{enumerate}
\usepackage{cleveref}

\newcommand{\quotes}[1]{``#1''}
\newcommand{\expt}[1]{\left \langle #1 \right \rangle}

\def \forall{\text{ for all }}

\definecolor{dark-gray}{gray}{0.3}

\crefname{equation}{Eq.}{Eqs.}

\newcommand{\paren}[1]{\left ( #1 \right )}
\newcommand{\braces}[1]{\left \{ #1 \right \} }
\newcommand{\abs}[1]{\left | #1 \right |}
\newcommand{\bracket}[1]{{\left [ #1 \right ]}}
\renewcommand{\min}[2]{\ensuremath{\text{min}\braces{ #1 ,\phantom{k} #2  }}}

\newcommand{\myequationmult}[1]{\begin{align} #1 \end{align}}

\newcommand{\myequation}[1]{\begin{equation}\begin{aligned} #1 \end{aligned} \end{equation}}
\newcommand{\myequationn}[1]{\begin{align*} #1 \end{align*}}

\newcommand{\thickhline}{\noalign{\hrule height 0.8pt}}
\newcommand{\appendref}[1]{({Append. \ref{#1})}}
\newcommand{\figref}[1]{{Fig. \ref{#1}}}

\def \temp{\ensuremath{T}\xspace}
\def \tamb{\ensuremath{T_{a}}\xspace}
\def \posit{\ensuremath{\vec{{r}} }\xspace}

\def \dens{\ensuremath{\rho}\xspace}
\def \densprime{\ensuremath{\alpha_{bee}}\xspace}
\def \mantdens{\ensuremath{\rho_{m}}\xspace}

\def \maxdens{\ensuremath{\rho_{max}}\xspace}
\def \thermotax{\ensuremath{\chi}\xspace}

\def \beepress{\ensuremath{P_{b}}\xspace}

\def \avclustrad {\ensuremath{{R}_{0}}\xspace}
\def \clustrad {\ensuremath{{R}}\xspace}
\def \dimlesssize{\ensuremath{\mathcal{V}}\xspace}
\def \beevol{\ensuremath{\dimlesssize_{0}}\xspace}

\def \press{\ensuremath{P}\xspace}
\def \metab{\ensuremath{\mathcal{M}}\xspace}
\def \cond{\ensuremath{{k}}\xspace}
\def \darcy{\ensuremath{{\kappa}}\xspace}
\def \airflow{\ensuremath{\vec{u}}\xspace}
\def \cels {\ensuremath{\text{C}}\xspace}
\def \degc {\ensuremath{^{\circ} \cels}}
\def \cair {\ensuremath{\mathcal{C}}\xspace}
\def \airvisc{\ensuremath{\eta}\xspace}

\def \airexpans{\ensuremath{\alpha_{air}}\xspace}

\def \airspecweight{\ensuremath{\gamma_{air}} \xspace}

\def \mycdot{\cdot \xspace}

\usepackage[nolabel]{fnbreak}
\usepackage{cancel}

\usepackage{soul}

\begin{document}
%\preprint{}

\title{
Collective thermoregulation in bee clusters  }
\author{Samuel A. Ocko}
\affiliation{Department of  Physics, Massachusetts Institute of Technology, Cambridge, Massachusetts 02139, USA}
\author{L. Mahadevan}
\affiliation{School of Engineering and Applied Sciences, Department of Physics, Harvard University, Cambridge, Massachusetts 02138, USA}
\affiliation{Wyss Institute for Biologically Inspired Engineering , Cambridge, Massachusetts 02138, USA}

\date{\today} 

\begin{abstract}
Swarming is an essential part of honeybee behaviour, wherein thousands of bees cling onto each other to form a dense cluster that may be exposed to the environment for several days. This cluster has the ability to maintain its core temperature actively without a central controller, and raises the question of how this is achieved. We suggest that the swarm cluster is akin to an active porous structure whose functional requirement is to adjust to outside conditions by varying its porosity to control its core temperature. Using a continuum model that takes the form of a set of advection-diffusion equations for heat transfer in a \emph{mobile} porous medium, we show that the equalization of an effective  \quotes{behavioural pressure}, which propagates information about the ambient temperature through variations in density, leads to effective thermoregulation.   Our model extends and generalizes previous models by focusing the question of mechanism on the form and role of the behavioural pressure, and allows us to explain the vertical asymmetry of the cluster (as a consequence of buoyancy driven flows), the ability of the cluster to overpack at low ambient temperatures without breaking up at high ambient temperatures, and the relative insensitivity to large variations in the ambient temperature. Finally, our theory makes testable hypotheses for how the cluster bee density should respond to externally imposed temperature inhomogeneities, and suggests strategies for biomimetic thermoregulation.

\hrulefill
\end{abstract}
%{\bf Keywords:} Swarms, Honeybees, Thermoregulation, Active Porous Media
%\pacs{03.67.Lx, 03.67.Pp, 05.30.Pr}
\maketitle

\section{Introduction}

%Introduce the problem, say what they do
{Honeybees are masters of cooperative thermoregulation, and indeed need to be able to do so for surviving during winter, raising their brood, frying predatory wasps, or during swarming. Swarming is an essential part of colony reproduction during which a fertilized queen leaves the colony with about 2,000-20,000 bees, which cling onto each other in a \emph{swarm cluster}, typically hanging on a tree branch, for times up to several days, while scouts search for a new hive location \cite{seeley2010honeybee}.} During this period, the swarm cluster regulates its temperature by forming a dense surface  \emph{mantle} that envelopes a more porous interior \emph{core}. At low ambient temperatures, the cluster contracts and the mantle densifies to conserve heat and maintain its internal temperature, while at high ambient temperatures the cluster expands and the mantle becomes less dense to prevent overheating in the core. {Over this period, when the swarm is in limbo before moving to a new home}  the cluster adjusts its shape and size to maintain the bees at a tolerable temperature, and is able to regulate the core temperature to within a few degrees of a homeostatic set point of $35\degc$ over a wide range of ambient conditions. 

%Self-organization problem
The swarm cluster is able to control its core temperature without a centralized controller to coordinate behaviour in the absence of any long-range communication between bees in different parts of the cluster \cite{Heinrich:1981ws}. Instead, the thermoregulation behaviour of a  swarm cluster emerges from the collective behaviour of thousands of bees\cite{Anderson:2002cq} who know only their local conditions. So how can bees in a cluster, each acting on very limited local information, control a core temperature that they are ignorant of? Early work on swarm clusters, and the related problem of winter clusters \cite{Stabentheiner:2003tx, heinrich1981insect}, used continuum models for variations in bee density, and temperature as determined by the diffusion of heat in a metabolically active material. Most of these models \cite{OmholtLonvik, LEMKE:1990ji, Omholt:1995ux, Eskov:2009ia} assumed that the bee know their location and the size of the cluster, contrary to experimental evidence. However, a new class of models initiated by Myerscough\cite{Myerscough:1993ue} are based on local information, as experimentally observed; these models are qualitatively consistent with the presence of a core and mantle, but are unable to maintain a high core temperature at low ambient temperatures. Further refinements of these models that account for bee thermotaxis and also use only local information\cite{Watmough:1995wn, SUMPTER:2000ez}, yield the observed mantle-core formation and good thermoregulation properties at low ambient temperatures, while allowing for an \emph{increased} core temperature at very low ambient temperatures, observed in some clusters\cite{SouthwickHypoHomeo, Fahrenholz:1989ky, Heinrich:1981jh, Heinrich:1981ws}. However, the thermotactic mechanism that defines these models of bee behaviour causes the cluster to break up at moderate to high ambient temperatures,  unlike what is observed. 

%Our model, what we do
Here we present a model for swarm cluster thermoregulation \cite{Note6} that results from the collective behaviour of bees acting based on local information, yet propagates information about ambient temperature throughout the cluster. Our model yields good thermoregulation and is consistent with experiments at both high and low temperatures, with a cluster radius, temperature profile, and density profile qualitatively similar to observations, without leading to cluster breakup. In Section 2, we  outline the basic principles and assumptions behind our model. In Section 3, we formulate our model mathematically, characterize the number of parameters in it, and solve the governing equations using a combination of analysis and numerical simulation in Section 4, and compare our results qualitatively with observations and experiments. We conclude with a discussion in Section 5, where we suggest a few experimental tests of our theory.

\section{Model assumptions}

Any viable mechanism for cluster thermoregulation {consistent with experimental observations} should have the following features: a) The behaviour of a cluster must result from the collective behaviour of bees acting on local information, not through a centralized control mechanism. b) The bee density of a cluster must form a stable mantle-core profile. c) The cluster must expand (contract) at high (low) ambient temperatures to maintain the maximum interior \quotes{core} temperature robustly over a range of ambient temperatures. 

\begin{figure}[h!] %  figure placement: here, top, bottom, or page
%            \label{fig:third}
\begin{center}
\includegraphics[width = .44 \textwidth]{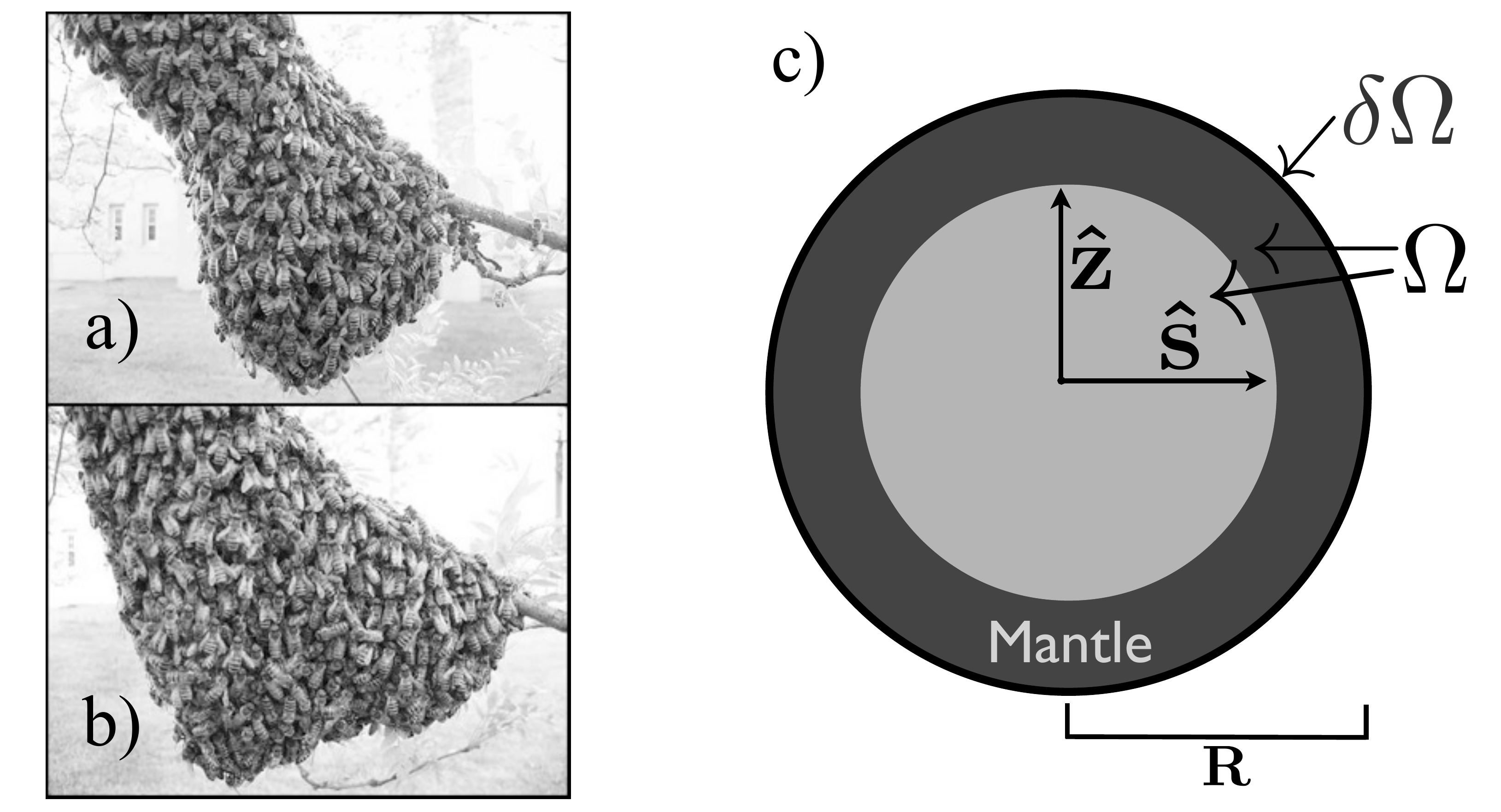}
\caption{A cluster at an ambient temperature of a) $11\degc$ and b) $27 \degc$, approximately 12 cm across(Photo courtesy of \cite{Cully:2004fv}). Note that number of bees inside the cluster is nearly constant; change in cluster width is a result of changes in bee packing fraction \cite{seeley2010honeybee}.
 c) Schematic of interior ($\Omega$), and boundary ($\delta \Omega$), with mantle-core structure. ${\hat{s}, \hat{z}}$ are radial, vertical directions in polar coordinates, \clustrad is the cluster radius. }
 \label{fig:Schematic}\
\end{center}
\end{figure}

This suggests that any quantitative model of the behaviour of swarm clusters requires knowledge of the transfer of heat through the cluster, the movement of bees within the cluster, and how these fields couple to each other. The basic assumptions and principles behind our model are as follows:
\begin{enumerate}
\item The only two independent fields are the packing fraction of bees in the cluster in the cluster \dens(1-porosity) and the air temperature \temp. These then determine the bee body temperature and metabolism which can be written as a function of the local air temperature, while convection of air in the form of upwards air currents depends entirely on the global bee packing fraction and temperature profiles.  
\item We treat the cluster as an active porous structure, with a packing fraction-dependent conductivity and permeability. Bees metabolically generate heat, which then diffuses away through conduction and is also drawn upwards through convection.  The boundary of the cluster has an air temperature that is simply the ambient temperature. 
\item
Cold bees prefer to huddle densely, while hot bees dislike being packed densely. In addition, bees attempt to push their way to higher temperatures. The movement of bees is determined by a behavioural variable which we denote by \quotes{behavioural pressure} \beepress(\dens, \temp), which we use to characterize their response to environmental variables such as local packing fraction and temperature. In terms of this variable, we assume that the bees move from high to low behavioural pressure, a notion that is similar in spirit to that of \quotes{social forces} used to model pedestrian movements \cite{SocialForces}. Here, we must emphasize that behavioural pressure is \emph{not} a physical pressure. For packing fraction to be unchanging, behavioural pressure must be constant throughout the cluster. 
\item
We assume that the number of bees in the cluster is fixed and that the cluster is axisymmetric with spherical boundaries, whose radius \clustrad is \emph{not} fixed (\figref{fig:Schematic}). At higher temperatures, the clusters become elongated and misshapen, so that the assumption is no longer accurate; however, this does not change our results for thermoregulation qualitatively. In general, to determine the shape of the cluster, we must account for \emph{both} heat and force balance, but we leave this question aside in the current study.
\end{enumerate}

A model based on these assumptions can be used to study both the equilibria and dynamics. However, systems which reach equilibrium quickly are best understood in terms of their \emph{equilibrium} behaviour rather than the specifics of \emph{how} this equilibrium is reached. Since a swarm cluster is a constantly changing network of attachments between bees, as bees grab onto and let go of nearby bees, and can also detach and reattach themselves at different points on the surface of the cluster, these \quotes{microscopic} dynamical processes allow the cluster to quickly equilibrate to changes in ambient conditions \cite{Heinrich:1981ws, Heinrich:1981jh}. Thus, we will  mostly focus on the resulting equilibria, but consider the slow dynamical modes that allow it to respond to large scale weak forcing, as they are particularly important in the context of cluster stability. %
%
%
%%%%%%%%%%%%%%%%%%%%%Start of Mathematical Formulation %%%%%%%%%%%%%%%%%%%%%%%
\section{Mathematical formulation}
The two independent fields in our model are bee packing fraction $\dens(\posit,t) \in [0,1]$, and the air temperature $\temp(\posit,t)$; while both of these are functions of space$(\posit)$ and time$(t)$, we will focus primarily on the equilibrium behaviour of these fields. 
For a static cluster of bees modeled as an active porous medium that generates heat and is permeable to air, the heat generated metabolically must balance heat lost due to conduction and convection, so that 
\myequation{
\dens \metab(\temp)  +\nabla \mycdot \paren{\cond(\dens) \nabla{\temp}} - \cair \airflow \mycdot \nabla \temp = 0 \mid_{\posit \in \Omega}\\ \label{eqn:advdiffus_units}
\temp = \tamb \mid_{\posit \in \delta \Omega}, 
}
where $\cond(\dens)$ is the packing fraction-dependent conductivity (Power/[Distance $\times$ Temperature]), \metab(\temp) is the metabolic heat production rate of the bees per unit volume, and \cair is the volumetric heat capacity of air(Energy/[Volume $\times$ Temperature]). We model the conductivity of the cluster as arising from a superposition of random convection currents within the cluster which are suppressed at high bee packing fraction and the bare conductivity of the bees treated as a solid, and approximate this by a function $\cond(\dens) = \cond_{0} \tfrac{1-\dens}{\dens}$. Although this form diverges as $\dens \to 0$ and random convection currents are unsuppressed, the $\dens$ never vanishes in the interior of the cluster, so that this limitation is not a problem. Likewise, by bounding \dens from above, we prevent $\cond$ from vanishing in the cluster. We further assume that the {mean flux per unit area} \airflow (Distance/Time) is determined by Darcy's law for the flow of an incompressible buoyant fluid through a porous medium, so that:
\myequationmult{
\airflow = \bracket{ \airspecweight  \airexpans  \paren{\temp - \tamb}  \hat{z} - \nabla \press}  {\darcy(\dens)}/{\airvisc}\mid_{\posit \in \Omega} \label{eqn:airflow_units}\\ 
 \nabla \mycdot \airflow = 0, \qquad \press = 0 \mid_{\posit \in \delta \Omega.} \label{eqn:press_units}
}

Here \cref{eqn:airflow_units} relates $\airflow$ to the effects of thermal buoyancy and the pressure gradient\cite{Note3}, while \cref{eqn:press_units} is just the incompressibility condition, with $\darcy(\dens)$ the packing fraction-dependent permeability(Distance$^{2}$), \airexpans is the coefficient of thermal expansion(Temperature$^{-1}$), \airspecweight is the specific weight of air(Pressure/Distance), \airvisc is the viscosity of air(Pressure $\times$ Time), and \press is air pressure.
We assume that the permeability of the cluster may be approximated via the Carman-Kozeny equation $\darcy(\dens) = \darcy_{0} \tfrac{\paren{1-\dens}^{3}}{{\dens^{2}}}$, used to describe the permeability of randomly packed spheres \cite{nield2006convection}.

In general, the bee metabolic activity is not a constant, and depends on a number of factors such as temperature, age, oxygen and carbon dioxide concentration etc.\cite{Heinrich:1981ws, Southwick:1985eb, NagyStalloneCO2}. To keep our model as simple as possible, we start by assuming that the metabolic rate is \emph{temperature-independent}, with $\metab(\temp) = \metab_{0}$ and show that this is sufficient to ensure robust thermoregulation, setting apart the details of the calculations that show that our model can also yield robust thermoregulation with a temperature-dependent metabolic rate \appendref{append:tempdepend}. 

To close our set of equations, we still must relate $\dens(\posit)$ to $\temp(\posit)$, which we do by making a hypothesis that bees respond to local packing fraction and temperature changes by changing their packing fraction to equalize an effective behavioural variable which we call the behavioural pressure. Our formulation of this behavioural pressure relies on two assumptions that are based on observations:
\begin{enumerate}
\item In their clustered state, bees have a natural packing fraction which is a function of the local temperature. This natural packing fraction decreases with increasing temperature, and increases with lower temperature, until it reaches a maximum packing fraction \maxdens. Effectively, cold bees prefer to be crowded, while hot bees dislike being crowded, consistent with a variety of experiments in the field and in the laboratory \cite{Heinrich:1981ws, Cully:2004fv, SouthwickHypoHomeo, heinrich1981insect}.
\item In addition to having a temperature-dependent natural packing fraction, bees also like to push their way towards higher temperatures. This will cause areas of equal \emph{local} temperature to pack more densely at low \emph{ambient} temperatures, consistent with observations\cite{Heinrich:1981ws}.
\end{enumerate} 

With these constraints in mind, a minimal model for bee behavioural pressure suggests the piecewise function:
\myequation{
\beepress(\dens, \temp)	 = \label{eqn:bee_press_units}
 \left\{
     \begin{array}{lr}
    	-  \thermotax  \temp + \abs{\dens -\mantdens(\temp) } &\dens \leq  \maxdens \\
	\infty							&\dens > \maxdens,
     \end{array}\right.     
}

where $\mantdens(\temp) =  \min{\maxdens}{\dens_{0} - \densprime \temp}$ is the natural packing fraction. The constant $\dens_{0}$ is dimensionless and represents the baseline for natural packing fraction; $\densprime$ has units of temperature$^{-1}$ describes how the natural packing fraction changes with temperature. \thermotax also has units of temperature$^{-1}$ and describes how bees push their way towards higher temperatures. Behavioural pressure becomes infinite at $\dens > \maxdens$ to enforce the maximum packing density. \cite{Note4} 

{We emphasize that our model assumes that \emph{individual} bees have no independent homeostatic set points for temperature or packing fraction; the behavioural pressure depends on a \emph{combination} of $\dens$ and $\temp$.} Furthermore, we note that the bee packing fraction in a cluster at equilibrium can never be lower than the natural packing fraction; in our formulation, the behavioural pressure is always higher at lower packing fractions and indeed is the cause of the basic aggregation behaviour that creates the cluster. Additionally, the absolute behavioural pressure is of no consequence; only gradients and relative values are important\cite{Note2}. Further evidence for a behavioural pressure comes from experiments \cite{Stabentheiner:2003tx} who observed a relation between the ambient temperature and the core density, even when the core temperature remained approximately constant.

To complete the formulation of our problem, we need to specify boundary conditions for the temperature and packing fraction fields. As we have stated earlier, the surface bees will be at the ambient temperature; furthermore, since they can freely expand and contract to minimize their behavioural pressure, we assume that they will be at their natural packing fraction $\mantdens(\tamb)$.

Comparing our model with earlier models such as the Myerscough model \cite{Myerscough:1993ue} and the Watmough-Camazine model \cite{Watmough:1995wn}, we note that both fit into this general behavioural pressure framework\appendref{append:PreviousWorkBeePress}. However, our work differs from these studies in that we synthesize and generalize the implicit relation between bee behaviour and their environmental variables in terms of the behavioural pressure, choosing a form for behavioural pressure which combines elements from both models and is consistent with experimental observations. In addition, we account for the role of fluid flow and convection of heat in the cluster, which breaks the vertical symmetry, and is potentially relevant in heat transport at high Rayleigh number 
${Ra} = \frac{\airspecweight \alpha  \cair}{\airvisc \cond}  \paren{\temp - \tamb} \darcy \clustrad$.  \\

\textbf{Dimensionless equations:}

To reduce the number of parameters in our model, we make our equations dimensionless. We note that the total number of bees in the cluster is constant; as we are using continuum model, this means that the total bee volume within the cluster $\iiint \dens dv$ is fixed. Rather than defining cluster size by number of bees, we define it by dimensionless total bee volume $\dimlesssize = \iiint \dens dv/ \beevol$, where \beevol is the total bee volume of a typical cluster, i.e. the average volume of a bee times the average number of bees in a cluster\appendref{append:est}. Upon setting a characteristic length scale to be the radius of a sphere of volume $\beevol$, $\avclustrad = \sqrt[3]{\frac{3 \beevol }{4 \pi}}$, we write the constraint on total bee volume as: $\iiint \dens dv= \paren{4\pi/3} \dimlesssize $. 
Scaling the temperature so that typical ambient temperature of $15\degc \to 0$, and the goal temperature of $35\degc \to 1$, we use the dimensionless variables: 
\def \mainbodydeltatemp{{20\degc}}
\myequationn{
   \temp \to \frac{\temp - 15 \degc}{\mainbodydeltatemp}, \quad &\tamb \to \frac{\tamb - 15 \degc}{\mainbodydeltatemp}  \\
\darcy_{0} \to \darcy_{0}   \frac{\airspecweight \airexpans  \cair}{\airvisc  \metab_{0} \avclustrad} 
 \paren{\mainbodydeltatemp}^{2},& \quad
\cond_{0} \to   \frac{\cond_{0} \paren{\mainbodydeltatemp}}{ \avclustrad^{2} \metab_{0}}\\
\metab_{0} \to 1, \qquad
\densprime \to \densprime  \paren{\mainbodydeltatemp},& \quad
\thermotax \to \thermotax  \paren{\mainbodydeltatemp}}

and write the dimensionless form of \cref{eqn:advdiffus_units,eqn:airflow_units,eqn:press_units,eqn:bee_press_units} as:
\myequationmult{
\dens    + \nabla \mycdot \paren{\cond(\dens) \nabla{\temp}} - \airflow \mycdot \nabla \temp = 0 \label{eqn:en_balance_dim_less} \\
\airflow = \bracket{  \paren{\temp - \tamb}  \hat{z} - \nabla \press}  \darcy(\dens), \qquad   \nabla \mycdot \airflow = 0. \label{eqn:air_flow_dim_less}\\
\cond(\dens) = \cond_{0}  \frac{1-\dens}{\dens},     \quad 
\darcy(\dens) = \darcy_{0}  \frac{\paren{1-\dens}^{3}}{\dens^{2}} \label{eqn:heat_functions_dim_less} \\
\temp = \tamb \mid_{\posit \in \delta \Omega}, \qquad \press =  0 \mid_{\posit \in \delta \Omega}\label{eqn:bound_values_dim_less} \\
\beepress(\dens, \temp)	 = 
 \left\{
     \begin{array}{lr}
    	\temp  \thermotax + \abs{\dens -\mantdens(\temp) } &\dens \leq  \maxdens \\
	\infty							&\dens > \maxdens,
     \end{array}\right.      \label{eqn:bee_press_dimless}     \\
     \mantdens(\temp) =  \min{\maxdens}{\dens_{0} - \densprime \temp}, \label{eqn:base_dens_dimliss}}
for the packing fraction and temperature profiles of the cluster which depend on the dimensionless parameters $\dens_{max}, \dens_{0}, \densprime, \thermotax, \tamb, \cond_{0}, \darcy_{0}, \dimlesssize $\cite{Note1}.

\textbf{Information transfer through equalization of behavioural pressure:}

\begin{figure}[h!] %  figure placement: here, top, bottom, or page
%   \centering
 \label{fig:beepressuresolving}            
            \includegraphics[width=0.45\textwidth]{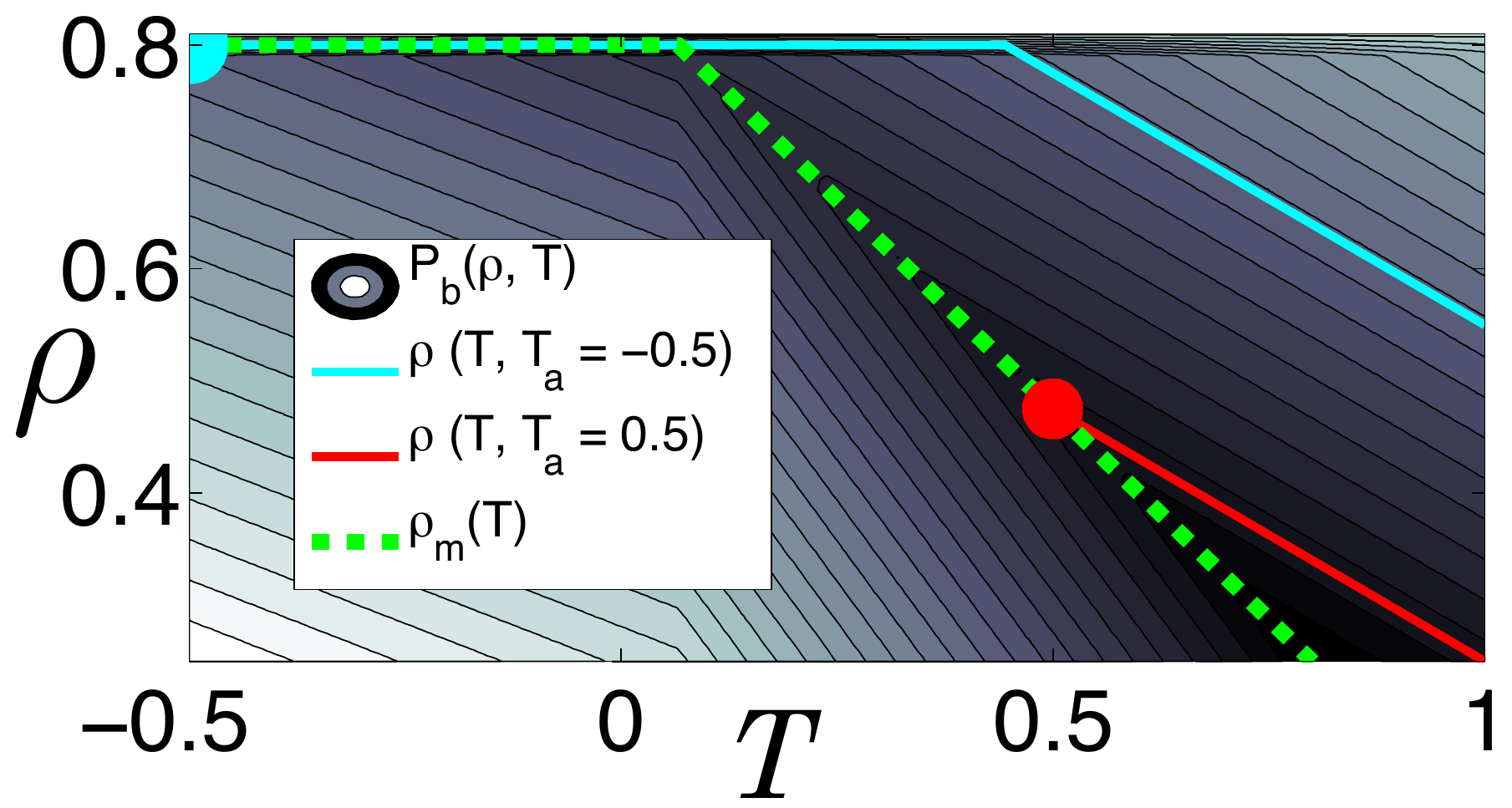} 
\caption{\footnotesize{Graphical representation of \cref{eqn:bee_press_dimless,eqn:base_dens_dimliss,eqn:bee_density_func}, showing how local packing fraction is obtained through behavioural pressure (Online version in colour). 
  {At low \tamb, bees on the mantle will pack at \dens = \maxdens(upper circle), and behavioural pressure is high throughout the cluster leading to higher interior packing fraction(upper solid line) and overpacking. At high \tamb, bees on the mantle will pack more loosely(lower circle), and low behavioural pressure throughout the cluster leads to low interior packing fraction(lower solid line). The solid lines run along contour lines of equal $\beepress(\dens, \temp)$, as behavioural pressure must be uniform through the cluster. Coefficients used are $\dens_{0} = 0.85, \densprime = 0.75, \maxdens = 0.8, \thermotax = 0.3$.}
}}            
\end{figure}

On the outer boundary of the mantle, the air temperature is equal to ambient temperature, so that minimizing the behavioural pressure requires that the packing fraction at the mantle will be the natural packing fraction at the ambient temperature $\mantdens(\tamb)$. At equilibrium, (4) then implies that the behavioural pressure in the mantle and thus throughout the cluster will be $-\tamb  \thermotax$. This means that throughout the cluster, we may write the local bee packing fraction as a function of both the local temperature and the ambient temperature, i.e $\beepress(\dens(\temp, \tamb), \temp) = -\tamb  \thermotax$ which we solve to find
\myequationn{
\dens(\temp, \tamb) = \min{\mantdens(\temp) + \thermotax  \paren{\temp - \tamb}}{ \maxdens}}
\myequation{
=\phantom{k} & \min{\dens_{0} + \temp \paren{\densprime + \thermotax} - \thermotax \tamb }{\maxdens}&\\ \label{eqn:bee_density_func}
=\phantom{k} &\min{ \dens_{0} + \temp  c_{0}+ \tamb   c_{1}}{\maxdens}, 
}
where we have made substitutions $c_{0} =  \densprime + \thermotax$, $c_{1} = - \thermotax$. Intuitively, the $c_{0}$ term characterizes the sensitivity of core temperature to ambient temperature, but the cluster cannot fully adapt at low $\tamb$ through this term alone; adaptation at lower ambient temperatures requires the $c_{1}$ term, which is eventually responsible for overheating in the core at very low $\tamb$. In {\figref{fig:beepressuresolving} we graph the packing fraction $\dens(\temp, \tamb)$ obtained by tracing contours of equal $\beepress$ which allows us to write the equilibrium local packing fraction everywhere in terms of the conditions at the boundary. Bees respond to their local conditions and move accordingly, and these variations in packing fraction propagate information about ambient temperature throughout the entire cluster without long-range communication.} Although we have mapped $\beepress(\dens, \temp) \to \dens(\temp, \tamb)$ for just one choice of behavioural pressure, our approach will work for any equation of state $\beepress(\dens, \temp)$ which  uniquely defines a stable packing fraction, i.e. with ${d \beepress}/{d\dens} >0$.

Now that we have obtained \cref{eqn:bee_density_func} from \cref{eqn:bee_press_dimless,eqn:base_dens_dimliss}, we put them aside and work with \cref{eqn:bee_density_func} directly. 
The set of equations(\ref{eqn:en_balance_dim_less} ,\ref{eqn:air_flow_dim_less} ,\ref{eqn:heat_functions_dim_less}, \ref{eqn:bound_values_dim_less}, and \ref{eqn:bee_density_func}) with the boundary condition that the surface temperature of the cluster is the ambient temperature completes the formulation of the problem to determine $\dens(\posit)$, $\temp(\posit)$.
\section{Simulations and results}

 While the boundary of the cluster is assumed to have spherical symmetry, the temperature and packing fraction fields inside do not have to have spherical symmetry owing to the effects of convection. However, the fields still have cylindrical symmetry, and therefore we can represent $\dens(\posit), \temp(\posit)$ as $\dens(s, z), \temp(s, z)$, where $s$ is the distance from the central axis and $z$ is the height. With this coordinate representation, we solve the governing equations (\ref{eqn:en_balance_dim_less} ,\ref{eqn:air_flow_dim_less} ,\ref{eqn:heat_functions_dim_less}, \ref{eqn:bound_values_dim_less}, and \ref{eqn:bee_density_func})  in a {spherically bounded domain using} a simple discretization scheme with $30$ values of $s$ and $60$ values of $z$ \appendref{app:finiteelement}. 
 
 Our choice of the dimensionless parameter  $ \cond_{0}=0.2 $ is constrained by experiments \cite{Southwick:1985eb}, while we estimate $ \darcy_{0} = 1$ though a simple calculation assuming the bees to be randomly packed spheres, and $\maxdens = 0.8$, slightly higher than the maximum packing fraction of spheres, as bees are more flexible. 
  \appendref{append:est}. 
  However, the parameters defining bee movement and behaviour, namely $c_{0}, c_{1}$, and $\dens_{0}$ are experimentally unknown. Guided by the  general observation that physiological performance is often improved by changing parameters while basic mechanisms remain unchanged, we optimize these parameters to achieve robust thermoregulation, i.e. the core temperature remains close to $1(35 \degc)$   over a \emph{range} of scaled ambient temperatures  $T_a \in [-0.7, 0.8]$ corresponding to a real ambient temperature $T_a \in [0,30]\degc$. For the choice of parameters $\dens_{0} = 0.85, c_{0} = 0.45, c_{1} = 0.3$, we find that within this wide range of $\tamb$ and a factor of three in $\dimlesssize$, the dimensionless core temperature stays in the range $0.7-1.3$ corresponding to a real temperature range of $~29 - ~41 \degc$, while the core temperature itself increases monotonically with $\dens_{0}$ in an analytically solvable way \appendref{append:DimAnal}.
 
\begin{figure}[h!] %  figure placement: here, top, bottom, or page
\label{fig:ConstMetabProfiles}
   \centering
%        \flushright   
      \includegraphics[width=0.45\textwidth]{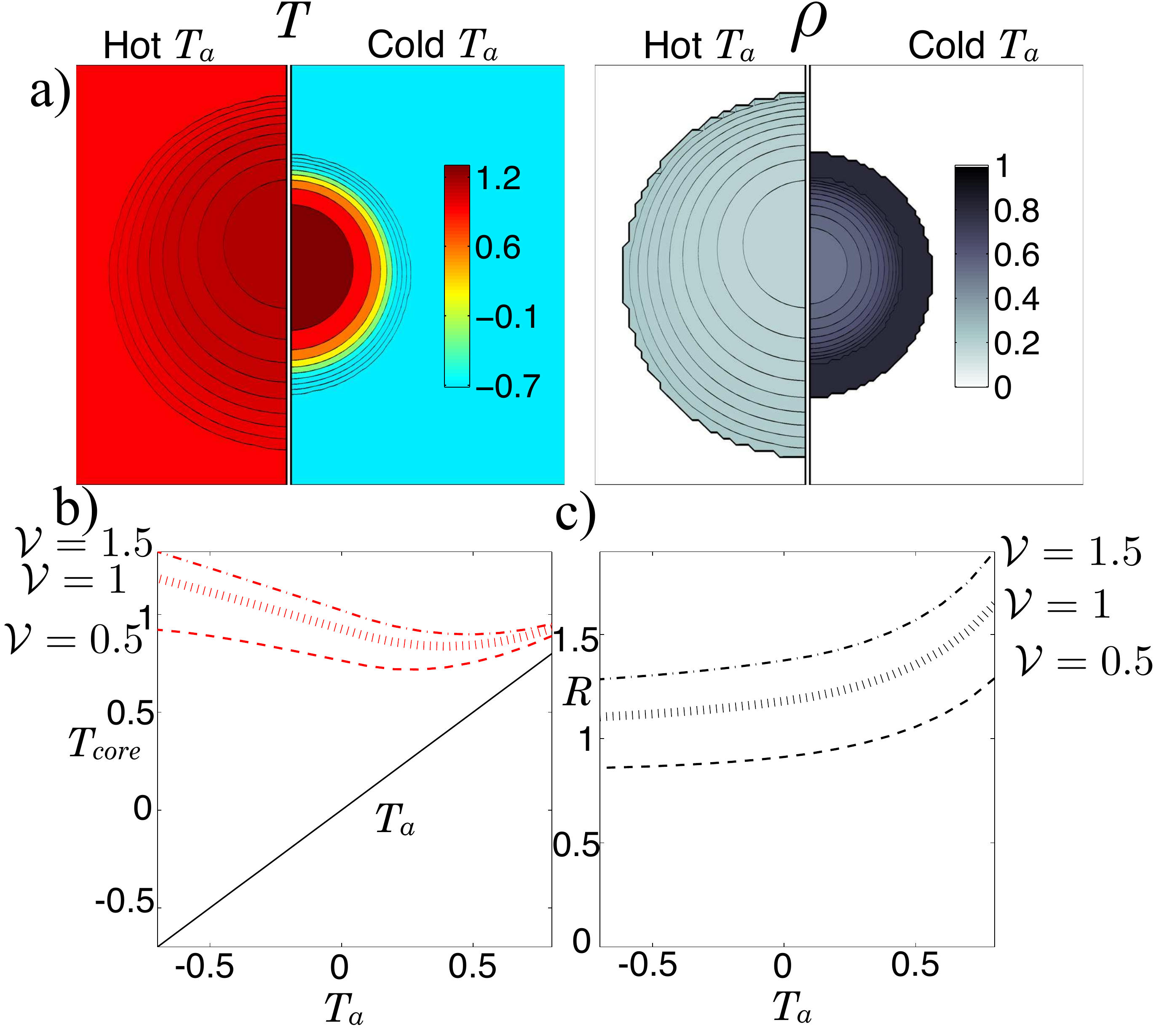}
\caption[justification = justified,singlelinecheck=false]{\footnotesize{Adaptation with temperature-independent metabolism(Online version in colour). a) Comparison of temperature and packing fraction profiles at high($0.8$) and low($-0.7$) ambient temperature,  {where the dimensionless total bee volume $\dimlesssize$ is $1$.} b) Core temperature as a function of ambient temperature and total bee volume shows that as $\dimlesssize$ increases the core temperature increases but adaptation {persists over a range of $\tamb$(also plotted to guide the eye.) } c) Cluster radius as a function of ambient temperature and total bee volume showing how clusters swell with temperature, consistent with experiment. For bee packing fraction, we choose coefficients of $\dens_{0} = 0.85, c_{0} = 0.45, c_{1} = 0.3, \maxdens = 0.8$. For heat transfer, we choose coefficients of $\cond_{0} = 0.2, \darcy_{0} = 1$. 
}}
\end{figure}

Our simulations also capture the qualitative mantle-core structure of the cluster (\figref{fig:ConstMetabProfiles})  with a dense mantle surrounding a sparse core. We also find that at high ambient temperatures, the cluster expands and the mantle thins, and at low ambient temperatures the cluster contracts and the mantle thickens. Furthermore, we find that the core temperature, which is also the maximum temperature of the cluster, is higher at \emph{low} ambient temperatures resulting from \quotes{overpacking}, consistent with experiments of \cite{SouthwickHypoHomeo, Fahrenholz:1989ky, Heinrich:1981jh, Heinrich:1981ws}, and predicted earlier \cite{Watmough:1995wn}. Finally, we note that the temperature profile is vertically asymmetric due to convection, causing the point of maximum temperature to rise above the geometric center of the cluster, as observed in experiments \cite{Heinrich:1981ws, Heinrich:1981jh}. 
\section{Discussion}

%1
We have shown that it is possible to provide a self-organized thermoregulation strategy in bee clusters over a range of observed ambient temperatures in terms of a few behavioural parameters. Our theory fits into a broader framework for understanding collective behaviour where the organism responds to the environment, but in doing so, changes it, and its behaviour until it reaches a steady state. Here, our model takes the form of a two-way coupling between bee behaviour and local temperature and packing fraction, quantified in terms of an effective behavioural pressure whose equalization suffices to regulate the core temperature of the cluster robustly. Although our choice of the form of the  behavioural pressure is likely too simplistic, it is consistent qualitatively with experimental observations, and we think it provides the correct framework within which we can start to quantify collective behaviour. Furthermore, our strategy might be useful in biomimetic settings. 
Our formulation for behavioural pressure shows that in a cluster, bee packing fraction should depend only on local temperature and the temperature of bees at the boundary, which effectively control the surface packing fraction. These dependencies might be measured by applying different temperatures to the surface and interior of an artificial swarm cluster. This observation provides an immediately testable prediction: a cluster may be \quotes{tricked} into overpacking and overheating its core by warming the bees just below the surface, while exposing the surface bees to a low temperature to increase behavioural pressure. As pointed out earlier, experiments and observations of bee core temperature and bee packing fraction \cite{Stabentheiner:2003tx} are consistent with these ideas, although a direct experiment of this type does not seem to have been carried out. Preliminary analysis of winter clusters shows that bee density can be written as a function of local temperature \cite{Note5}. 

%2
Our model adjusts well in response to changes in ambient temperatures, but it doesn't have the same level of tolerance to different total bee volume that honeybees exhibit. We have used a continuum model where the bees on the surface  exposed to the ambient temperature equilibrate to that temperature, but in reality the first layer of bees is hotter than ambient temperature and supports a large temperature gradient driven by heat from interior bees. This means that the surface bees feel an average temperature higher than ambient due to the interior bees, and and we predict this should give a large cluster a lower behavioural pressure than a small cluster at the same ambient temperature. This should reduce the sensitivity of core temperature to total bee volume, and we have confirmed this in simulations \appendref{append:SurfaceGradients}.

%3
We close with a description of some possible extensions of our study. Currently, our model ignores changes in shape of the cluster associated with force balance via the role of gravity, and the associated effects on thermoregulation.  In reality, the cluster is a network of connections between bees which changes shape and size due to a combination of mechanical forces and heat balance, and a complete theory must couple these two effects as well. 

Our heat balance equation and estimates\appendref{append:est} suggest that while convective terms are responsible for the asymmetry in the temperature profile, they do not play an important global role in thermoregulation. However, our calculation assumes a uniform packing which does not accurately represent the microscopic structure of the cluster, and thus we may have underestimated the Rayleigh number and importance of convection. At high ambient temperatures, swarm clusters are observed to \quotes{channelize}, where channels open up to ventilate. Studying a simple \quotes{behavioural pressure-taxis} dynamical law\appendref{append:StabResults} , we  find no linear instability that leads to channelization without an \quotes{anemotaxis} mechanism, where behavioural pressure increases with $|\airflow|$; instead we see only only one kind of linear instability which comes from the mathematical necessity of fixing cluster radius. This raises the question of whether there are anemotaxis mechanisms. If not, how can channelization result from bee-level dynamics and mechanics? We have also neglected active cooling, which includes fanning, evaporative cooling(which can give up to $\sim 50\degc$ of cooling), diffusion of heat through diffusion of bees, and effects of oxygen and CO$_{2}$ \cite{Southwick:1985eb, SouthwickVentilation, Hypoxia, NagyStalloneCO2}.  Finally, we have also neglect any implications of bee age distribution, despite knowledge of the fact that younger bees tend to prefer the core and produce less heat, while older bees prefer the mantle and can produce more heat\cite{Fahrenholz:1989ky, Heinrich:1981ws, Cully:2004fv}. Accounting for these additional effects will allow us to better characterize the ecological and possibly even evolutionary aspects of thermoregulation.

Thermoregulation is a necessity for a wide variety of organisms. When achieved collectively, individuals expend effort at a cost that accrues a collective benefit. The extreme relatedness of worker bees in a cluster and near-inability to reproduce implies that the difference between the individual and the collective is nearly nonexistent, so that cost and benefit are equally shared. However, many other organisms are faced with the \quotes{huddler's dilemma}\cite{haig2010huddler};  expending individual metabolic effort is costly, and benefits a group that is only partially related. Because genetic relatedness, metabolic costs, individual temperature and spatial positions are all easily measurable\cite{Gilbert:2009dg}, a collective thermoregulatory system is an ideal context in which to study the tangible evolution of cooperation and competition by building on our framework theoretically to account for competition and cooperation.

\section{Acknowledgments}
We thank James Makinson, Madeline Beekman, and Mary Myerscough for helpful discussions and sharing videos of swarm clusters, and Anton Stabentheiner for helpful discussions and sharing data on winter clusters. 
We thank an anonymous referee for very helpful and detailed suggestions on the manuscript. 
We thank Brian Lee and Zhiyan Wei for helpful suggestions, and the Henry W. Kendall physics fellowship (S.O), and the MacArthur Foundation (L.M.) for financial support.  
\onecolumngrid

%\center{Table of quantities and associated units}
\def \beeflow{\ensuremath{\vec{\text{J}}} \xspace}
 %Some more definitions
\def \kJ{\ensuremath{\text{kJ}} \xspace}
\def \joule{\ensuremath{\text{J}} \xspace}
\def \kg{\ensuremath{\text{kg}} \xspace}
\def \gram{\ensuremath{\text{gram}} \xspace}
\def \cm{\ensuremath{\text{cm}} \xspace}
\def \cmcube{\ensuremath{\text{cm}^{3} } \xspace}
\def \ml{\ensuremath{\text{mL}} \xspace}
\def \watt{\ensuremath{\text{W}} \xspace}
\def \hour {\ensuremath{\text{hour}} \xspace}
\def \minute {\ensuremath{\text{min}} \xspace}

\def \meter{\ensuremath{\text{m}} \xspace}
\def \sec{\ensuremath{\text{s}} \xspace}

\begin{center}
\begin{table}[H]
\caption{Table of quantities and associated units}
  \begin{tabular}{| c | c | c | c |}
    \hline				
    Quantity 							& Symbol 		& Units 							&Typical Values								\\ \thickhline	
    Bee Packing Fraction					& \dens 		& Dimensionless		    			&$\sim 0.2 - 0.8$ 								\\     \hline
    Bee Metabolic Rate					& \metab 		& Power/Volume 					&$\sim0.001 - 0.01 \watt/\cmcube$ \cite{Heinrich:1981jh, Southwick:1985eb}					\\     \hline 
    Heat Conductivity 						& \cond 		& Power/(Distance$\times$Temperature) &$\sim 0.0004 -0.006  \watt/(\cm \degc)$? \appendref{append:est} 				\\     \hline    
    Permeability 							& \darcy 		& Distance$^{2}$ 					& $(0.05 \cm)^{2}$?	\appendref{append:est}		\\    \hline
%   Flow related stuff
    Darcy Velocity						& \airflow 		& Distance/Time 					&$\sim 1 \cm/\sec$?  \appendref{append:est}		\\    \hline
    Air Heat Capacity						& \cair 		& Energy/(Volume$\times$Temperature)		&$1.2 \times 10^{-3} \joule/(\ml \  \degc$)				\\    \hline %http://www.engineeringtoolbox.com/air-properties-d_156.html

    Air Specific Weight 					& \airspecweight& Pressure/Distance				&$1.2 \times g/(\cm^{2} \ \sec^{2}$)				\\    \hline %http://www.engineeringtoolbox.com/air-properties-d_156.html
    
%    Gravity	 							& \grav 		& Distance/time$^{2}$ 				&$981  \cm/\sec^{2}$							\\    \hline%google
%    Air Density 							& \airdens		& Mass/Volume  					&$1.2 \times 10^{-3} \gram/\ml$						\\    \hline %http://www.engineeringtoolbox.com/air-properties-d_156.html

    Coefficient of thermal expansion 			& \airexpans		& Temperature$^{-1}$ 				&$1/300\degc$									\\    \hline%PV = NRT%       
    Air Viscosity 							& \airvisc 		& Pressure$\times$Time					&$1.8 \times 10^{-4} \gram/(\cm \ \sec)$				\\    \hline%1.8 × 10^(-5) Pascal*second /(gram /(cm * second)) + wikipedia
%Bee related stuff    
   \temp dependence of $\dens_{m}$			& \densprime	& Temperature$^{-1}$		   			&$\sim 0.04 \degc^{-1}?$ 											\\    \hline
    Behavioural Pressure					& \beepress	& Model Dependent \appendref{app:behavpressureunits}		   			&Unknown \appendref{app:behavpressureunits}			\\    \hline    
    Thermotactic Coefficient				& \thermotax	& Behavioural Pressure/Temperature 			&Unknown\appendref{app:behavpressureunits}					\\        
     \hline
    \end{tabular}
    \end{table}
\end{center}
\twocolumngrid

%
%\bibliography{OneForall}  

%
\clearpage
\newpage
\newpage
\newpage
\pagebreak

\onecolumngrid

%%%%%%%%%%%%%%%%%%%%%%%%%%%%%%%%%START OF FIRST BIT%%%%%%%%%%%%%%%%%%%%%%%%%%%%%%%%%%%%%%%%%%%%%%%%%

\appendix

\section{Units, parameter estimation, and dimensional analysis}

%\newpage
\subsection{Estimation of heat conductivity, metabolic rate, and permeability}\label{append:est}

For the metabolic rate and heat conductivity we use estimates from the experiments of Southwick \cite{Southwick:1985eb}, where a cluster of 4250 bees(608 grams) is put into a set of roughly planar, parallel honeycombs, and the temperature profile and oxygen consumption are measured. The bees are roughly uniformly distributed with a bee packing fraction $\dens$ of about $0.5$, for which the metabolic rate is roughly uniform, and the temperature is well approximated with a parabolic profile, with a temperature of $34\degc$ at the core, and $11\degc$ at the edge $9.5 \cm$ from the core, parallel to the combs. The combs insulate well, so heat transfer occurs primarily in the two directions parallel to the combs. Then,  within the cluster:
\myequationn{
\temp \approx 34 \degc - 23 \degc \paren{ \frac{x^{2} + y^{2}}{9.5 \cm^{2}}},\\
\nabla^{2} \temp \approx -4 \times  \frac{23 \degc}{\paren{9.5 \cm ^{2}}} \approx \frac{1.0 \degc}{\cm ^{2}}.
}
The oxygen consumption rate is measured to be $6.5  \ml/\minute$ which gives a volumetric metabolism of $0.0035  \watt/\cm^{3}$, assuming a bee specific weight of 1 $\gram/\cm^{3}$, and an oxygen to energy conversion of $3.5 \frac{\ml \ O_{2}}{\kg  \minute} \equiv  0.0012 \frac{\watt}{\gram}$. This metabolic rate agrees well with the experiments of Heinrich \cite{Heinrich:1981ws}. We now have all the pieces to calculate the conductivity using the conduction heat balance:
\myequationn{
\cond \nabla^{2} \temp  +  \dens \metab = 0 \Longrightarrow \cond =  \frac{1.7 \times 10^{-3} \watt}{\cm \degc}.
}
At $\dens = 0.8$, the maximum packing fraction we allow, the conductivity becomes close to the value of $2.4 \times 10^{-3} \watt/(\cm  \degc)$ for fur and feathers \cite{TempAndLife} as suggested by Southwick which gives us some level of confidence in the functional form $\cond_{0} \frac{1 - \dens}{\dens}$ we have chosen for conductivity. \\

To estimate $\darcy_{0}$, we use the Carman-Kozeny equation \cite{nield2006convection}. The average bee weighs about 0.14 grams, which corresponds to a sphere of diameter $\sim 0.65 \cm$. In the absence of any detailed information about the bee structure in the cluster, we treat the cluster as a system of randomly packed spheres. It would be interesting to measure and better understand convective gas and heat transfer within swarm clusters. Using this diameter in the Carman-Kozeny equation, we find:%
\myequationn{
\darcy_{0} = \frac{\text{D}^{2}}{180} \approx \paren{0.05 \cm}^{2}.
}
A typical cluster has about 10,000 bees, which is about 1.4 kg, giving $\clustrad_{0} = 7 \cm$. Plugging these values in\appendref{append:DimAnal}, we find dimensionless   conductivity and permeability to be:
\myequationn{
\cond_{0} \approx 0.2, \qquad
\darcy_{0} \approx 1.
}

\def \mydeltatemp {\ensuremath{35\degc - 15\degc}}

\subsection{Dimensional analysis}\label{append:DimAnal}

Our conditions for heat balance imply that:
\myequationn{
\dens \metab(\temp)  +\nabla \mycdot \paren{\cond(\dens) \nabla{\temp}} - \cair \airflow \mycdot \nabla \temp = 0 \mid_{\posit \in \Omega}\\
\temp = \tamb \mid_{\posit \in \delta \Omega}\\
\airflow = \bracket{ \airspecweight  \airexpans  \paren{\temp - \tamb}  \hat{z} - \nabla \press}  {\darcy(\dens)}/{\airvisc}\mid_{\posit \in \Omega} \\
 \nabla \mycdot \airflow = 0, \qquad \press = 0 \mid_{\posit \in \delta \Omega.}
}

while our equation for behavioural pressure reads:
\myequationn{
\beepress(\dens, \temp)	 = 
 \left\{
     \begin{array}{lr}
    	-  \thermotax  \temp + \abs{\dens -\mantdens(\temp) } &\dens \leq  \maxdens \\
	\infty							&\dens > \maxdens,
     \end{array}\right.     \\
\mantdens(\temp) =  \min{\maxdens}{\dens_{0} - \densprime \temp}     
}
We set our unit of length  to be $\clustrad_{0}$, so that the volume constraint becomes  $\iiint \dens dv= \paren{4\pi/3} \dimlesssize $. We make the transformation $\nabla \to \nabla/\clustrad_{0}$. Then our heat balance equations read:

\myequationn{
\dens \metab(\temp)  +\frac{1}{\clustrad_{0}^{2}}\nabla \mycdot \paren{\cond(\dens) \nabla{\temp}} - \frac{1}{\clustrad_{0}}\cair \airflow \mycdot \nabla \temp = 0 \mid_{\posit \in \Omega}\\
\temp = \tamb \mid_{\posit \in \delta \Omega}\\
\airflow = \bracket{ \airspecweight  \airexpans  \paren{\temp - \tamb}  \hat{z} -  \frac{1}{\clustrad_{0}}\nabla \press} {\darcy(\dens)}/{\airvisc}\mid_{\posit \in \Omega} \\
 \nabla \mycdot \airflow = 0, \qquad \press = 0 \mid_{\posit \in \delta \Omega.}
}

On making the substitutions:

\myequationn{
\temp \to \frac{\temp - 15\degc}{\mydeltatemp}, \qquad \tamb \to \frac{\temp - 15\degc}{\mydeltatemp}
}
leads to the goal temperature of $35\degc $ yielding a dimensionless temperature of unity, and a typical ambient temperature of $15\degc$ corresponding to a dimensionless temperature that vanishes. Then our system of equations becomes:
\myequationn{
\dens \metab(\temp)  +\frac{\mydeltatemp}{\clustrad_{0}^{2}}\nabla \mycdot \paren{\cond(\dens) \nabla{\temp}} - \frac{\mydeltatemp}{\clustrad_{0}} \cair \airflow \mycdot \nabla \temp = 0 \mid_{\posit \in \Omega}\\
\temp = \tamb \mid_{\posit \in \delta \Omega}\\
\airflow = \bracket{ \paren{\mydeltatemp}  \airspecweight \airexpans    \paren{\temp - \tamb}  \hat{z} -  \frac{1}{\clustrad_{0}}\nabla \press}  {\darcy(\dens)}/{\airvisc}\mid_{\posit \in \Omega} \\
 \nabla \mycdot \airflow = 0, \qquad \press = 0 \mid_{\posit \in \delta \Omega.}
}
\myequationn{
\beepress(\dens, \temp)	 = 
 \left\{
     \begin{array}{lr}
    	-  \thermotax \paren{\mydeltatemp} \temp + \abs{\dens -\mantdens(\temp) } &\dens \leq  \maxdens \\
	\infty							&\dens > \maxdens,
     \end{array}\right.     \\
\mantdens(\temp) =  \min{\maxdens}{\bracket{\dens_{0} - \densprime  15\degc} - \densprime \temp \bracket{\mydeltatemp}},
}
 We divide all terms in the heat equation by the base metabolism $\metab_{0}$ to yield:
\myequationn{
\dens \metab(\temp)/\metab_{0}  +\frac{\mydeltatemp}{\metab_{0} \clustrad_{0}^{2}}\nabla \mycdot \paren{\cond(\dens) \nabla{\temp}} - \frac{\mydeltatemp}{\metab_{0} \clustrad_{0}}  \paren{ \cair \airflow \mycdot \nabla \temp} = 0 \mid_{\posit \in \Omega}\\
\temp = \tamb \mid_{\posit \in \delta \Omega}\\
\airflow = \bracket{ \paren{\mydeltatemp} \airspecweight \airexpans    \paren{\temp - \tamb}  \hat{z} -  \frac{1}{\clustrad_{0}}\nabla \press}  {\darcy(\dens)}/{\airvisc}\mid_{\posit \in \Omega} \\
 \nabla \mycdot \airflow = 0, \qquad \press = 0 \mid_{\posit \in \delta \Omega.}.
}

Making the substitution 
\myequationn{\airflow \to  \airflow   \times  \frac{\mydeltatemp}{\metab_{0}\clustrad_{0}}  \cair }
%\myequationn{
%\dens \metab(\temp)/\metab_{0}  +\frac{\mydeltatemp}{\metab_{0} \clustrad_{0}^{2}}\nabla \mycdot \paren{\cond(\dens) \nabla{\temp}} - \airflow \mycdot \nabla \temp = 0 \mid_{\posit \in \Omega}\\
%%
%\temp = \tamb \mid_{\posit \in \delta \Omega}\\
%%
%\airflow =  \frac{\mydeltatemp}{\metab_{0} \clustrad_{0}} \mycdot \cair  \mycdot \bracket{ \paren{\mydeltatemp} \airspecweight \airexpans    \paren{\temp - \tamb}  \hat{z} -  \frac{1}{\clustrad_{0}}\nabla \press} \mycdot {\darcy(\dens)}/{\airvisc}\mid_{\posit \in \Omega} \\
%%
% \nabla \mycdot \airflow = 0, \qquad \press = 0 \mid_{\posit \in \delta \Omega.}.
%}
%
and the substitution 
\myequationn{
\press \to \frac{\press}{\paren{\mydeltatemp}  \airspecweight \airexpans    \clustrad_{0}}, 
}
leads to the set of equations:

\myequationn{
\dens \metab(\temp)/\metab_{0}  +\frac{\mydeltatemp}{\metab_{0} \clustrad_{0}^{2}}\nabla \mycdot \paren{\cond(\dens) \nabla{\temp}} - \airflow \mycdot \nabla \temp = 0 \mid_{\posit \in \Omega}\\
\temp = \tamb \mid_{\posit \in \delta \Omega}\\
\airflow =  \bracket{   \paren{\temp - \tamb}  \hat{z} - \nabla \press}    \frac{\paren{\mydeltatemp}^{2} \airspecweight \airexpans     \cair}{\metab_{0} \clustrad_{0} \airvisc}    {\darcy(\dens)} \mid_{\posit \in \Omega} \\
 \nabla \mycdot \airflow = 0, \qquad \press = 0 \mid_{\posit \in \delta \Omega.}.
}
Finally, making the substitutions for the coefficients 
\myequationn{
\metab \to \metab/\metab_{0}, \qquad \cond \to \cond  \frac{\mydeltatemp}{\metab_{0} \clustrad_{0}^{2}} \\
\darcy \to  \darcy \frac{\paren{\mydeltatemp}^{2} \airspecweight \airexpans     \cair}{\metab_{0} \clustrad_{0} \airvisc} ,
}

\myequationn{
\dens_{0} \to \dens_{0} - \densprime \paren{15\degc}, \qquad \densprime \to \densprime  \paren{\mydeltatemp}, \qquad \thermotax \to \thermotax  \paren{\mydeltatemp}.
}
leads to our full dimensionless set of equations for heat balance:
\myequationn{
\dens \metab(\temp)  + \nabla \mycdot \paren{\cond(\dens) \nabla{\temp}} - \airflow \mycdot \nabla \temp = 0 \mid_{\posit \in \Omega}\\
\temp = \tamb \mid_{\posit \in \delta \Omega}\\
\airflow =  \bracket{   \paren{\temp - \tamb}  \hat{z} - \nabla \press}    {\darcy(\dens)} \mid_{\posit \in \Omega} \\
 \nabla \mycdot \airflow = 0, \qquad \press = 0 \mid_{\posit \in \delta \Omega.}.}
while the dimensionless behavioural pressure reads: 
 \myequationn{
\beepress(\dens, \temp)	 = 
 \left\{
     \begin{array}{lr}
    	-  \thermotax  + \abs{\dens -\mantdens(\temp) } &\dens \leq  \maxdens \\
	\infty							&\dens > \maxdens,
     \end{array}\right.     \\
\mantdens(\temp) =  \min{\maxdens}{\dens_{0} -  \densprime \temp }, 
}
Our model has seven parameters $\dens_{max}, \dens_{0}, \densprime, \thermotax, \tamb, \cond_{0}, \darcy_{0}$, with an additional parameter for the total bee volume \dimlesssize, which varies from cluster to cluster.

\def \tempzero{\ensuremath{\temp_{\text{packed}}}\xspace}
\def \tempone{\ensuremath{\temp_{\text{empty}}}\xspace}
\def \otherdeltatemp {\ensuremath{\tempone  - \tempzero}}

\subsubsection{Note on further dimensional analysis} 

We note that, when the metabolic rate is temperature independent, the goal temperature and the typical independent ambient temperature have no bearing on the actual behaviour of he model, only on whether it represents effective thermoregulation. Then, we may set the temperature where the packing becomes maximally dense, $\tempzero = \frac{\dens_{0} - \maxdens}{\densprime}$, to be zero, and the temperature at which the packing fraction becomes zero, $\tempone = \frac{\dens_{0}}{\densprime}$, to be 1. We can then make slightly different substitutions for the coefficients:
\myequationn{
\metab \to 1, \qquad \cond \to \cond  \frac{\otherdeltatemp}{\metab_{0} \clustrad_{0}^{2}} \\
\darcy \to  \darcy \frac{\paren{\otherdeltatemp}^{2} \airspecweight \airexpans     \cair}{\metab_{0} \clustrad_{0} \airvisc} ,
}

\myequationn{
\dens_{0} \to \maxdens, \qquad \densprime \to \maxdens, \qquad \thermotax \to \thermotax  \paren{\otherdeltatemp}.
}

We may remove the parameter \dimlesssize by setting the length scale to be the fully packed radius of this particula \emph{particular} cluster rather than the fully packed radius of a \emph{typical} cluster. Doing this makes the total bee volume constraint become  $\iiint \dens dv= \paren{4\pi/3} $, and requires the substitutions:
\myequationn{
\cond \to \frac{\cond}{ \dimlesssize^{2/3}}, \qquad
\darcy \to \frac{\darcy}{ \dimlesssize^{1/3}}.
}
We now find that there are now only five free parameters $\dens_{max}, \thermotax, \tamb, \cond_{0}, \darcy_{0}$. We do not carry out this extra analysis in the main body of the paper because this causes us to lose sight of what the goal core temperature, typical ambient temperatures, and typical cluster sizes are.

\subsection{Units of behavioural pressure}\label{app:behavpressureunits}
Our model is based on behavioural pressure being uniform at equilibrium. The units and typical values of behavioural pressure are unknown, as  any sets of dynamical equations for bee movement will result in the same equilibrium where behavioural pressure, whose units depend on the set of dynamical equations used, remains constant. For example, a simple taxis model is one where:
\myequationn{
\frac{d \dens}{dt} = - \nabla \mycdot \beeflow,  \quad \beeflow = -\nabla \beepress
}
would mean that behavioural pressure has units of Distance$^{2}$/Time. A more complicated evaporation/condensation model would have the form

\def \transf{\ensuremath{\mathcal{T}} \xspace}
\myequationn{
\frac{d \dens(\posit)}{dt} = \iiint \bracket{\paren{\beepress(\posit') - \beepress(\posit)} \frac{e^{\frac{|\posit - \posit'|^{2}}{2 \sigma^{2}}}}{\sigma^{3}}  \transf(\dens(\posit), \dens(\posit'))} d^{3} \posit',
}
where $\sigma$ is the evaporation and condensation radius and $\transf$ is some transfer coefficient would give units of 1/Time. A yet more complicated model involving mechanical compressibility or viscosity would have yet another set of dimensions for behavioural pressure. All of these models would, however, yield the same static solution.

\section{Behavioural pressure formalism and its antecedents in previous models}\label{append:PreviousWorkBeePress}

To understand how our behavioral pressure formalism fits in with previous models, we compare them within this framework. We note that previous modes have defined density in terms of $bees/\cm^{3}$ instead of packing fraction; to go between the two, bees may be assumed to have water density, and a packing fraction of 1 corresponds to
 $(1 \text{ gram})/m_{bee}$  $\text{bees}/\cm^{3}$.

 The Myerscough model assumes that the bee density depends only on \emph{local} temperature, and thus can be written as $\beepress(\dens, \temp) = |\dens - \mantdens|$, where $ \mantdens = 8 \ \text{bees}/\text{cm}^{3}  \paren{1 - \frac{\temp}{40 \degc}}$.   The Watmough-Camazine model defines a dynamical law for the bee density via the equations:
 \myequationn{
\dot{\dens} = -\nabla \mycdot \beeflow\\
\beeflow = -\mu(\dens) \nabla \dens - \dens \chi(\temp) \nabla \temp, 
 }
where $\mu(\dens) > 0$ is a motility function, and $\chi(\temp)$ is a thermotactic function. This may be written as 

\myequationn{
\beeflow = -\dens  \bracket{\frac{\mu(\dens)}{\dens} \nabla \dens + \chi(\temp) \nabla \temp} = -\dens \nabla \beepress, 
}
where the behavioural pressure $\beepress$ is defined as:
\myequationn{
\beepress(\dens, \temp) = \underbrace{\int {\frac{\mu(\dens')}{\dens'} d \dens' }}_{\text{Density Component}} +  \underbrace{\int {\chi\paren{\temp'} d \temp'.}}_{\text{Temperature Component}}
}

We note that as $\mu(\dens) >0 \forall \dens$, the density component is minimized as $\dens \to 0$, and the density at the surface must be fixed to prevent the cluster from falling apart, unlike in our behavioral pressure framework. We believe the Watmough-Camazine behavioural pressure allows the packing fraction at the mantle to become too high. 
Additionally, because the behavioural pressure can be divided into temperature and density components, the point of highest density will always be at the same temperature, regardless of size or ambient temperature, which is not observed experimentally. This is in contrast with our formalism.

%%%%%%%%%%%%%%%%%%%%%%%%%%%%%%%%%END OF FIRST BIT%%%%%%%%%%%%%%%%%%%%%%%%%%%%%%%%%%%%%%%%%%%%%%%%%
%
%
%%%%%%%%%%Method of our numerical technique%%%%%%%%%%%%%

\section{Numerical methods}
\subsection{Method of solving for equilibrium}
To reach equilibrium, we use an iterative scheme, described by the following pseudocode:

\def \ratio{ \mathcal{R}}
\def \coeff {\textbf{c}}
%%%%%%%%%%%%%%%%
\begin{algorithmic}
\State $\temp(\posit) \gets  \tamb$
\State $\clustrad \gets \clustrad_{0}/\sqrt[3]{\maxdens}$
\State $\dens(\posit) \gets \left\{
     \begin{array}{lr}
    	\maxdens &:|\posit| \leq \clustrad  \\
	0		&:|\posit| > \clustrad  ,\\		
     \end{array}
     \right. $
     \State 
\Repeat
	\State \emph{Solve for \airflow at fixed temperature and density, then solve for \temp at fixed \metab and \airflow}
	\State Find $\airflow, \press$ such that: $\airflow = \bracket{  \paren{\temp - \tamb}  \hat{z} - \nabla \press}  \darcy(\dens)$, $\nabla \mycdot \airflow = 0$
	\State Find $\temp_{new}(\posit)$ such that: $\dens \metab(\temp) = -\nabla \mycdot \paren{\cond(\dens) \nabla{\temp_{new}}} - \airflow \mycdot \nabla \temp_{new}$
	\State $\temp(\posit) \gets \temp(\posit) + \paren{\temp_{new}(\posit) - \temp(\posit)}$
	\State
	%Update Density
	\State $\dens_{new}(\posit) \gets \dens(\temp(\posit), \tamb)$
	\State $\dens(\posit) \gets \dens(\posit) + \coeff_{\dens}\paren{\dens_{new}(\posit) - \dens(\posit)}$
	\State
	\State \emph{Expand or contract the bee packing fraction and temperature profiles to normalize total bee volume}
	\State $\ratio \gets \sqrt[3]{\frac{\dimlesssize}{\iiint \dens}}$ \emph{Scaling ratio}
	\State $\clustrad \gets \clustrad   \ratio$
	\State $\dens(\posit) \gets \dens(\posit/\ratio)$
	\State $\temp(\posit) \gets \temp(\posit/\ratio)$
\Until{converged}
\end{algorithmic}

The intermediate steps can be solved as a system of linear equations. Note that this method does not add or remove cells to vary cluster radius and conserve the total number of bees; it grows and shrinks a fixed number of cells. The solution is considered to be converged when $\dens_{new} = \dens$, $\temp_{new} = \temp$, $\ratio = 1$ to within $10^{-10}$, which takes about $100-200$ iterations, about a minute on a laptop. All simulations were done using MATLAB.

%%%%%%%%%%%%%%%%
%\section{Solving for equilibrium profile}
%Can't be whitespace in \bib command

\def \ij{\ensuremath{ij} \xspace}
\def \ijp{\ensuremath{i'j'} \xspace}
\def \ijij{\ensuremath{ij, i'j'} \xspace}
\def \grain{\ensuremath{n} \xspace}
\def \cellw{\ensuremath{w} \xspace}

\def \discflow{{u}}
 \newcommand{\harmean}[2]{H \paren{ #1, #2 }}
 \newcommand{\harmeanb}[2]{H \bracket{ #1, #2 }}

%\newpage
\subsection{Discretization of space}\label{app:finiteelement}

To solve for the temperature and density profiles, we must first discretize the system. While spherical symmetry is broken due to convection, the system retains rotational symmetry about its axis. We therefore use cylindrical coordinates, where each cell is given indices $(i, j)$, and has coordinates which represent the distance from the central axis $s_{ij}$ and the vertical coordinate $z_{\ij}$, where 
\myequationn{
s_{\ij} = \paren{i + \frac{1}{2}} \frac{\clustrad}{\grain}, \qquad
z_{\ij} = \paren{j-\frac{1}{2}} \frac{\clustrad}{\grain}.
}
\grain is the radius of the cluster in cells. All cells with $s_{\ij}^{2} + z_{\ij}^{2} \leq \clustrad$ are in the interior of the cluster, while all cells with $s_{\ij}^{2} + z_{\ij}^{2} > \clustrad$  are at the exterior of the cluster, subject the the boundary conditions $\temp_{\ij} = \tamb$, $\press_{\ij} = 0$, $\dens_{ij} = 0$. 

The volume of each cell with coordinates $(i, j)$ is $V_{\ij} = 2 \pi \cellw^{2} s_{\ij}$, where $\cellw = \frac{\clustrad}{\grain}$ is the width of each cell. Each cell (i, j) neighbors four other cells, $(i, j+1), (i, j-1), (i+1, j), (i-1, j)$, with the exception of cells which border the axis $(i = 0)$, which only have three neighbors. The area shared by cell $(i, j)$ and its outside horizontal neighbor $(i+1, j)$ is $2 \pi \cellw \paren{s_{\ij} + \frac{\cellw}{2}}$. The area shared by cell $(i, j)$ and its vertical neighbor $(i, j+1)$ is $2 \pi \cellw s_{\ij}$.
 \begin{figure}[htbp] %  figure placement: here, top, bottom, or page
    \label{fig:graining}
    \centering
    \includegraphics[width = 1.23in]{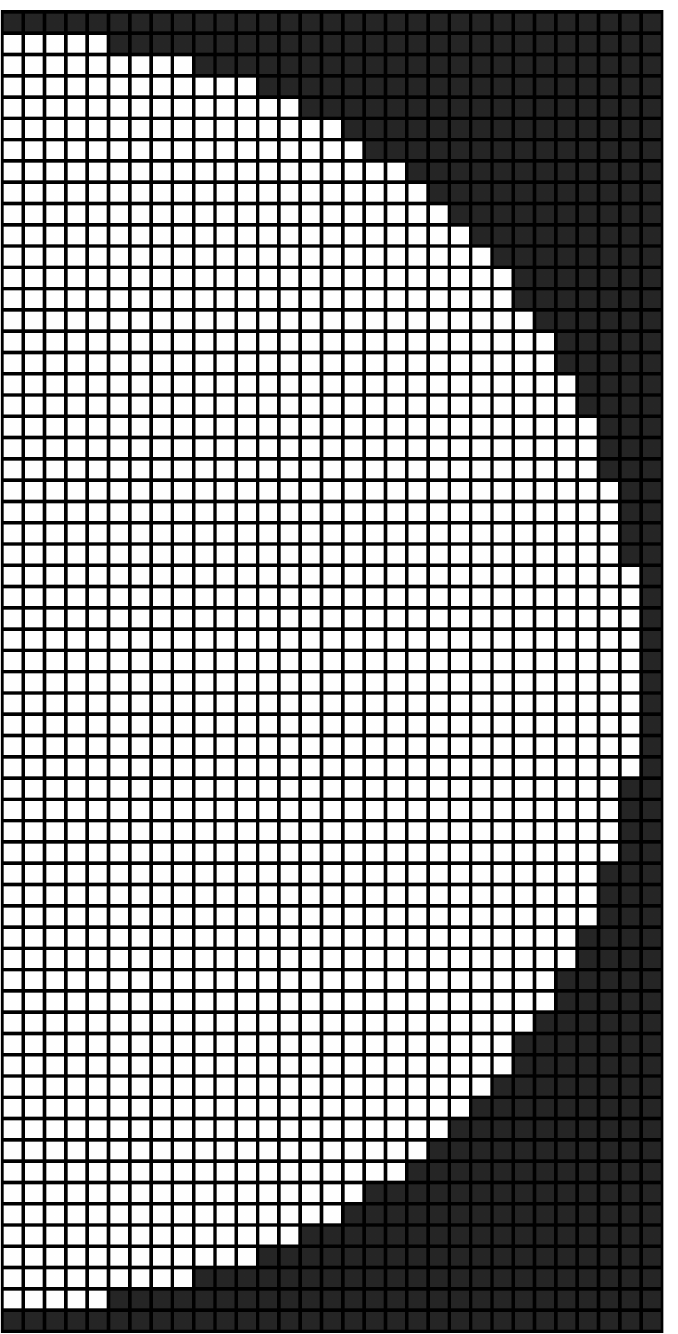} 
    \caption{Cell layout of a cluster $30$ cells in radius. Interior cells are coloured white, exterior cells are coloured grey.}
 \end{figure}
\subsubsection{Heat equations}
For heat balance, we discretize our (now dimensionless) heat equations as 
\myequationn{\frac{dQ_{\ij}}{dt} =
\underbrace{ \metab(\temp_{\ij}) \dens_{\ij} V_{\ij} }_{\text{Metabolism}}
%Conductivity
-\underbrace{ \sum_{\expt{\ijp}} \frac{A_{\ijij}}{\cellw}   \harmean{\cond(\dens_{\ij})}{ \cond(\dens_{\ijp})} \paren{\temp_{\ijp} - \temp_{\ij} } }_{\text{conduction}}
%Convection
+ \underbrace{ \sum_{\expt{\ijp}}{ \discflow_{\ijij}  \bracket{-\theta  \paren{ \discflow_{\ijij}  }\temp_{\ij} + \theta  \paren{- \discflow_{\ijij}  }\temp_{\ijp} }}}_{\text{convection}} = 0.
}
The first term corresponds to metabolic heat generation, and the second term is heat conduction, where $\sum_{\expt{\ijp}}$ is the sum of all $(i', j')$ neighboring $(i, j)$, with the heat conductance between two neighboring cells depending on the harmonic mean of the conductance of each cell; $\harmean{a}{b} = \frac{2}{{1}/{a} + {1}/{b}}$. 
 The third term represents convective heat transfer, with $\discflow_{\ijij}$  the air flow from cell $(i, j)$ to $(i', j')$(Units of dimensionless volume/time due to discretization). 
 
 We have chosen an \quotes{upwinding} scheme \cite{Patankar:269851}; when air flows \emph{out} of a cell, the outwards heat flux is determined by the temperature of \emph{that} cell. When air flows \emph{into} a cell from a neighboring cell, the inwards heat flux is determined by the temperature of the \emph{neighboring} cell where the air originates. This scheme uses the Heaviside step function:
 \myequationn{
\theta (x) =  \left\{
     \begin{array}{lr}
    	0 & x \leq 0  \\
	1& x > 0.\\		
     \end{array}
     \right. 
}
\subsubsection{Solving for buoyancy driven flow}
The air flow from cell $\ij$ to neighboring cell $\ijp$ is given by:
\myequationn{\discflow_{\ijij} = \harmean{\darcy(\dens_{\ij})}{\darcy(\dens_{\ijp})}  \frac{A_{\ijij}}{\cellw} 
 \bracket{ \paren{\press_{\ij} - \press_{\ijp}}   + \paren{z_{\ijp}-z_{\ij}} \paren{\frac{\temp_{\ij} + \temp_{\ijp}}{2} - \tamb}}
}
where the air conductance between two cells again depends on the harmonic mean of the permeability of each cell, and the pressure is set so that air flow is conserved in every cell i.e. ${ \sum_{\expt{\ijp}}{ \discflow_{\ijij}}} = 0$ for all cells $(i, j)$. This yields a set of linear equations that can be solved easily.

%%%%%%%%%%%%%%%%%%%%%%%END OF SECOND BIT%%%%%%%%%%%%%%%%%%%%%%%%%%%%%%%

%\newpage
%%%%%%%%Temperature Dependent Metabolic Rate %%%%%%%%%%%%%%%%%%%

\section{Temperature dependent metabolic rate}\label{append:tempdepend}

To formulate our temperature-dependent metabolic rate, we note that bees have a higher base metabolic rate at high temperatures than at low temperatures. At moderate temperatures, bees on the mantle keep their body temperatures approximately $3 \degc$ above ambient temperature, and at below $15\degc$, they will \quotes{shiver} to keep their body temperature at  $18\degc$\cite{Heinrich:1981jh}. Assuming a constant coefficient of thermal transfer between a bee and the surrounding air, this leads to a formulation of metabolism by shivering at air temperatures below $15 \degc$, and gives us the full piecewise function for metabolic rate. Our metabolic rate for high temperatures comes from experiments involving oxygen consumption in swarm clusters \cite{Heinrich:1981jh}.
\myequationn{
\notag
\metab(\temp)= \metab_{0}  \left\{
     \begin{array}{lr}
    	1 + \frac{15\degc - \temp}{3\degc} &:\temp < 15\degc \\
	1 +  \frac{\temp - 15\degc}{10\degc}						&: \temp \geq 15\degc ,\\		
     \end{array}
     \right. 
}
where the base metabolism, $\metab_{0}$ has units of power/volume. 

\begin{figure}[h!]
   \centering
%            \label{fig:third}
            \includegraphics[width=0.30\textwidth]{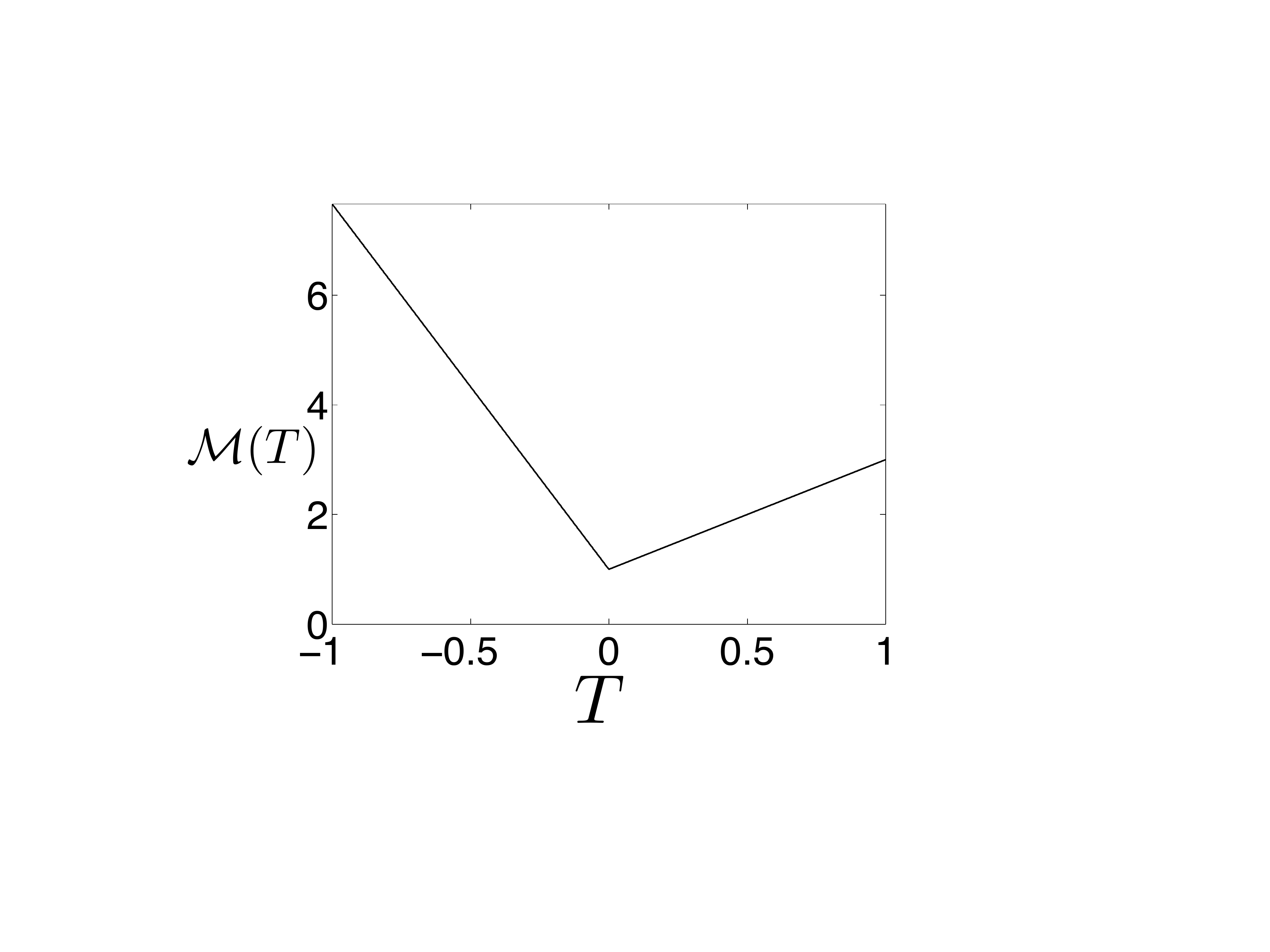}
\caption{Metabolic rate as a function of temperature.}
\end{figure}
\begin{figure}[h!] %  figure placement: here, top, bottom, or page
\label{fig:ChangingMetabProfiles}
   \centering
      \includegraphics[width=0.45\textwidth]{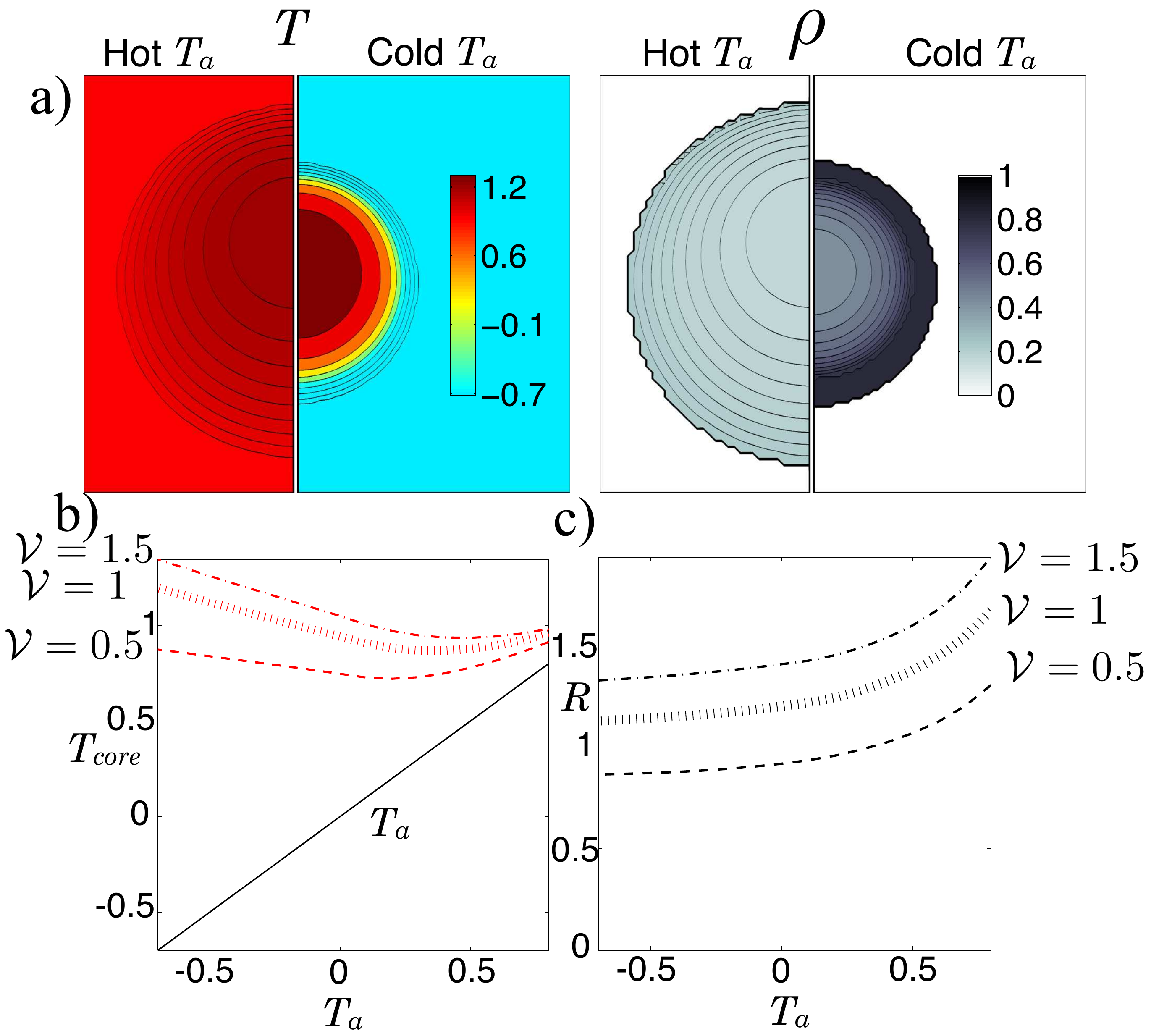}   
\caption{Adaptation with temperature-dependent metabolic rate (Online version in colour). Note that cluster behavior is nearly the same as for constant metabolism(Compare to \figref{fig:ConstMetabProfiles}). 
 a) Comparison of temperature and packing fraction profiles at high($0.8$) and low($-0.7$) ambient temperature, where the dimensionless total bee volume $\dimlesssize$ is $1$. b) Adaptation is nearly the same as for temperature independent metabolic rate. $\tamb$ is also plotted to guide the eye. c) Cluster radius is also nearly the same as for temperature independent metabolic rate. 
For packing fraction, we choose coefficients of $\dens_{0} = 0.85, c_{0} = 0.5, c_{1} = 0.25, \maxdens = 0.8$. 
}
\end{figure}

The simulated cluster with temperature-dependent metabolic rate has the same qualitative features as when we use temperature-independent metabolic rates.  We set $\metab_{0}$ to be   $0.00175  \watt/\cm^{3}$, half as high as when \metab was set temperature-independent, because now it represents the \emph{minimal} metabolic rate, not the \emph{uniform} one  (\figref{fig:ChangingMetabProfiles}). Applying the same dimensional analysis, this results in the values of $\cond_{0}, \darcy_{0}$ being doubled, giving $\cond_{0} = 0.4$, $\darcy_{0} = 2.$

%\clearpage

%%%%%%%%%%%%%%%%%%%%%%%END OF THIRD BIT%%%%%%%%%%%%%%%%%%%%%%%%%%%%%%%

\section{Accounting for the role of finite bee size on thermoregulation}\label{append:SurfaceGradients}

\def \beelength {\textbf{L}_{\text{bee}}}

In the paper, we have defined the temperature at the boundary to be the ambient temperature, and we assumed the surface bees feel the ambient temperature, which sets the behavioural pressure accordingly. In reality, bees point their heads inwards, and feel a temperature gradient driven by the heat produced by interior bees. Therefore, it may be more realistic that the behavioural pressure is set by the temperature a slight distance inwards from the surface. If we include the effects of convection, this implies that spherical symmetry must be broken, and the temperature becomes not just a function of distance, but also dependent on angle. Therefore, to close our set of equations without having to delve deeper into the question of cluster shape, we must neglect upwards convection and only consider conduction. This gives us the system of equations:

\myequationn{
\dens \metab_{0}    + \nabla \mycdot \paren{\cond(\dens) \nabla{\temp}}= 0 , \qquad
\cond(\dens) = \cond_{0}  \frac{1-\dens}{\dens},     \quad \\
\temp = \tamb \mid_{\posit \in \delta \Omega}, 
}
where the behavioural pressure is now set by the temperature a distance of $\beelength$ inside the cluster rather than by the ambient temperature:
\myequation{
=\phantom{k} &\min{ \dens_{0} + \temp  c_{0}+ \temp(\clustrad - \beelength)   c_{1}}{\maxdens}.
}
Assuming  $\beelength =1$ cm, slightly shorter than the body length of a worker bee, for an average cluster radius of $7$ cm we find that the dimensionless $\beelength$ of $\approx 0.14$. We simulate the system for dimensionless total bee volumes of $0.5, 1,$ and $3$, with the same parameters of $c_{0} = 0.45$, $c_{1} = 0.3$, $\maxdens = 0.8$ as were used in paper. In the first system, to test thermoregulation with this effect, we choose $\beelength = 0.14$, and compensate for the slightly lower average behavioural pressure by increasing $\dens_{0}$ to 0.95. In the second system, we test thermoregulation without this effect, and so we choose $\beelength = 0, \dens_{0} = 0.85$, as we did in the main body. We find that for large clusters, the temperature gradient created by the interior bees lowers behavioural pressure and loosens the cluster. This mitigates overheating in the core and sensitivity to total bee volume, e.g. at $\tamb = -0.7$, $\temp_{core}$ varies $\sim 50\%$ more with cluster size when behavioural pressure is set by ambient temperature rather than the temperature beneath the surface. We note that in the models of Myerscough \cite{Myerscough:1993ue} and Watmough-Camazine\cite{Watmough:1995wn}, a similar shielding of surface bees from ambient air and reduction of sensitivity to cluster size was achieved through use of a heat transfer coefficient, with units of Power/(Area Temperature) between the cluster surface and ambient air.
\begin{figure}[h!] %  figure placement: here, top, bottom, or page
%            \label{fig:third}
\begin{center}
      \includegraphics[width=0.70\textwidth]{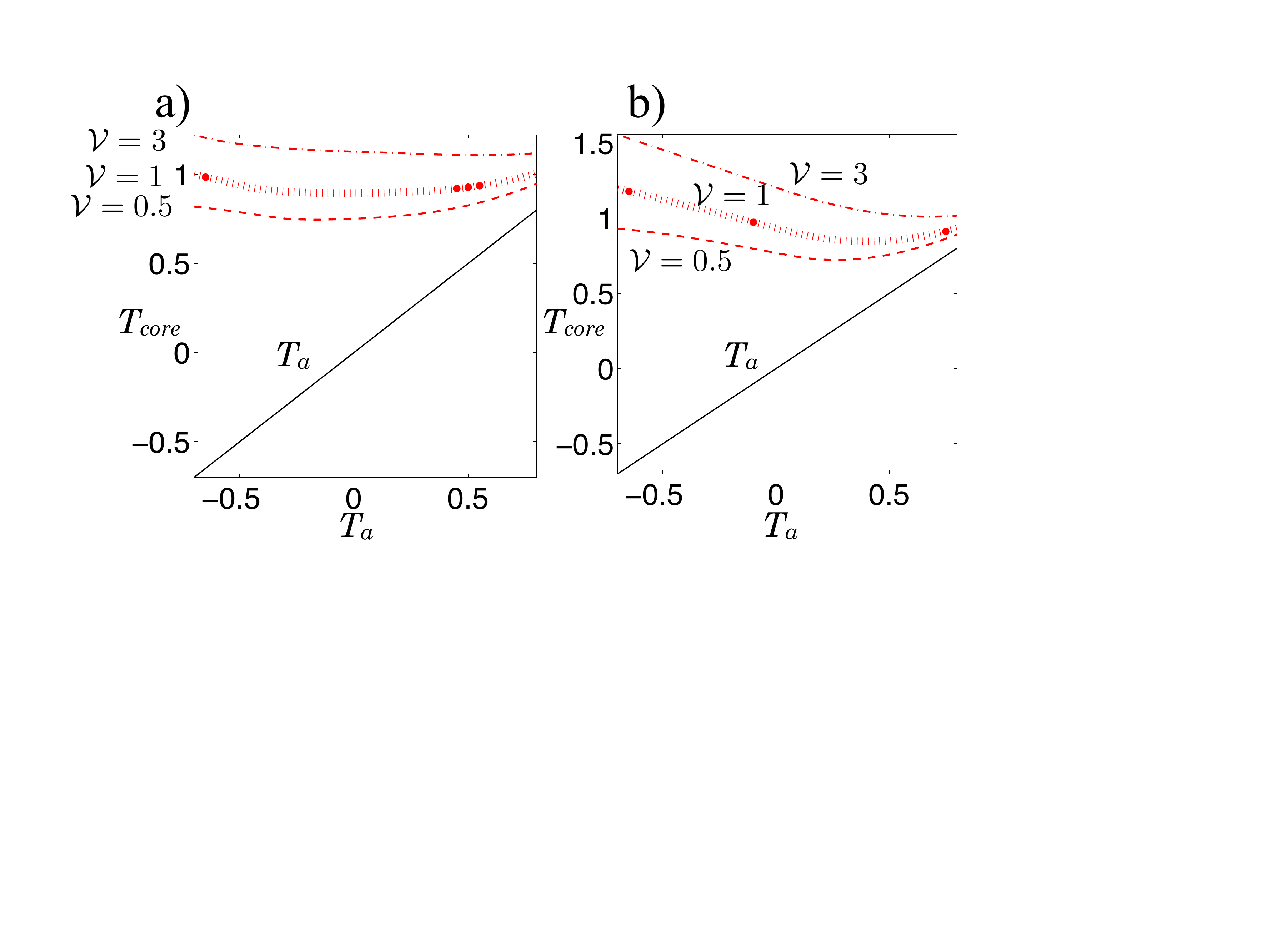}
%\subfigure{
%            \includegraphics[height=0.3\textwidth]{Runs/SurfaceGradientEffect/CoreTempsWithEffect}            
%}
%\subfigure{
%            \includegraphics[height=0.3\textwidth]{Runs/SurfaceGradientEffect/CoreTempsWithoutEffect}            
%}
\caption{Comparison of core temperatures(Online version in colour).  a) Behavioural pressure is set by the temperature slightly below the surface of the cluster, and sensitivity to total bee volume is mitigated. b) Behavioural pressure is set by ambient temperature, and sensitivity of core temperature to total bee volume is considerably higher. $\tamb$ is also plotted in each to guide the eye. Note that figure b) is very similar to Fig. \ref{fig:ConstMetabProfiles}b), as it results from the same equations except for convection.}
\end{center}
\end{figure}

%%%%%%%%%%%%%%%%%%%%%%%END OF GRADIENT BIT%%%%%%%%%%%%%%%%%%%%%%%%%%%%%%%

\def \stabmat{\ensuremath{\textbf{M}} \xspace}
\def \wavenum{\ensuremath{\textbf{k}_{\phi}} \xspace}
\def \MyDelt{\ensuremath{\Delta} \xspace}
\def \mydelt{\ensuremath{\delta} \xspace}
\def \myphase{\ensuremath{e^{i \wavenum \phi}} \xspace}
\def \myinvphase{\ensuremath{e^{-i \wavenum \phi}} \xspace}

\section{Results of linear stability analysis}\label{append:StabResults}
Solving for linear stability using simple \quotes{behavioural pressure-taxis} dynamics \appendref{append:StabMethods}, we find that all clusters simulated at a temperature-independent metabolic rate are stable. However,  at low ambient temperatures, clusters with a temperature-dependent \appendref{append:tempdepend} metabolic rate can be linearly unstable via an overheating instability, which we believe to be a relic of fixing the boundaries of the cluster. In this instability, bees from the core move to the mantle, increasing the mantle thickness and insulation. This causes the core to heat up, increasing its behavioural pressure causing even more bees to move from the core to the mantle, leading to eventual runaway. We only see this instability when using a temperature-dependent metabolic rate, where an increased core temperature results in a greater net metabolic rate, aggravating the problem. 
\begin{figure}[h!] %  figure placement: here, top, bottom, or page
%            \label{fig:third}
\begin{center}
\label{fig:InstabEig}
\includegraphics[width = .95 \textwidth]{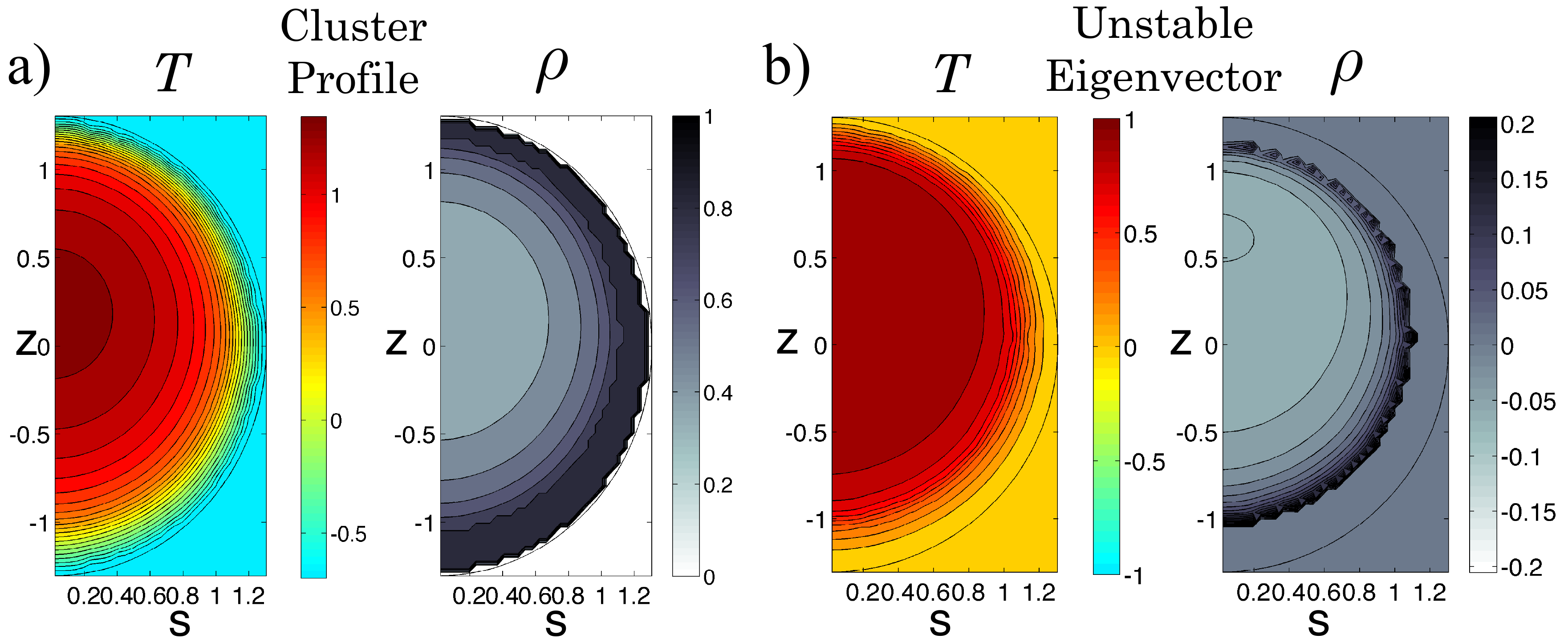}
%\subfigure[Cluster Profile]{
%            \includegraphics[height=0.3\textwidth]{Runs/InstabGraphs/AmbT-0p70}            
%}
%\subfigure[unstable Eigenvector of Linear Response Matrix]{
%            \includegraphics[height=0.3\textwidth]{Runs/InstabGraphs/InstabilityOutput}            
%}
\caption{a) Cluster profile.  b)Contour plots of unstable eigenvector of linear response matrix(Online version in colour). Circle represents boundaries of the cluster, and a temperature-dependent metabolic rate is is used. $\beeflow_{0}= 1, \dimlesssize = 1.5, \tamb = -0.7$.  $\dens_{0} = 0.85, c_{0} = 0.5, c_{1} = 0.25, \maxdens = 0.8$. For heat transfer, we choose coefficients of $\cond_{0} = 0.4, \darcy_{0} = 2$. Note that the packing fraction very close to the boundary is fixed because $\dens = \maxdens$.
}
\end{center}
\end{figure}

Dynamical behavior  which allows the cluster radius to vary would suppress this instability, as bees moving from the core to the mantle would result in an expansion of the cluster, increasing the surface area and cooling the core. However, a dynamical model which allows the boundaries of the cluster to change requires a better understanding of the bee-level structure and the mechanics within a swarm cluster, and we leave this aside here.

%\newpage
%%%%%%%%Linear Stability %%%%%%%%%%%%%%%%%%%
\subsection{Methods for calculating linear stability}\label{append:StabMethods}
Having solved for the equilibrium state, we want to find if this state is stable or unstable. To do so, we must define some dynamical laws for bee movement. Choosing a simple \quotes{behavioural pressure-taxis} behaviour, where $\beeflow \propto -\dens \nabla \beepress$, $\frac{d \dens}{dt} = -\nabla  \mycdot \beeflow$ gives us the complete set of dynamical equations: 
\myequationn{
\dens \dot{\temp}   = \dens \metab(\temp)  +\nabla \mycdot  \paren{\cond(\dens) \nabla{\temp}} - \cair \airflow \mycdot \nabla \temp = 0 \mid_{\posit \in \Omega} = \dens F\bracket{\dens, \temp}, \qquad
\temp = \tamb \mid_{\posit \in \delta \Omega}\\
\airflow = \bracket{ \airspecweight  \airexpans  \paren{\temp - \tamb}  \hat{z} - \nabla \press}  {\darcy(\dens)}/{\airvisc}\mid_{\posit \in \Omega}, \qquad
 \nabla \mycdot \airflow = 0, \qquad \press = 0 \mid_{\posit \in \delta \Omega.}\\
\beeflow = - \beeflow_{0}  \dens \nabla \beepress, \qquad \beeflow = 0 \mid_{\posit \in \delta \Omega}, \qquad \frac{d \dens}{dt} = -\nabla \mycdot \beeflow \mid_{\posit \in \Omega}  = G\bracket{\dens, \temp}
}
Here we vary $\beeflow_{0}$ over a wide range to reflect the large variations in bee movement and temperature time scales, and  define $F, G$ to be the functionals that determine the dynamics of the system. We also emphasize some issues about these choices: a) Our form assumes a substrate that the bees move on. In reality, in a swarm cluster the bees \emph{are} the substrate, and changes on one side of the cluster propagate mechanically to other parts of the cluster at a rate faster than the taxis rate. b) Bee movement is not necessarily a local movement down pressure gradients. Bees can disconnect from the cluster and reattach at a different point, which doesn't fit into the local gradient picture. c) There is a discontinuity in behavioural pressure and packing fraction at the boundary of the cluster, so this taxis model does not explain how the boundary of the cluster can change. Within these limits then, this choice of behaviour gives us a full set of differential equations. To determine if the equilibrium state of the system is linearly stable, we must determine whether the linear response matrix:
\myequationn{
\stabmat =    \begin{pmatrix} % or pmatrix or bmatrix or Bmatrix or ...
      \frac{d F}{d \temp} &  \frac{d F}{d \dens} \\
      \frac{d G}{d \temp} &  \frac{d G}{d \dens} \\
   \end{pmatrix}
}
has positive eigenvalues. We note that the equilibrium temperature and bee packing fraction profile we solved for is symmetric with respect to rotation about the central axis($\phi$ direction), and therefore we may partition the space of perturbations into subspaces defined by wave number \wavenum: $\MyDelt \temp(s, z, \phi) = \myphase \mydelt \temp(s, z) $, $\MyDelt \dens(s, z, \phi) = \myphase \mydelt \dens(s, z)$. These temperature and bee packing fraction perturbations will in turn give a change in airflow, pressure, and behavioural pressure: $\MyDelt \airflow(s, z, \phi) = \myphase \mydelt \airflow(s, z) $, $\MyDelt \press(s, z, \phi) = \myphase \mydelt \press(s, z) $, $\MyDelt \beepress(s, z, \phi) = \myphase \mydelt \beepress(s, z)$. All of these will change the temperature and bee packing fraction time derivatives which will be proportional to $\myphase$. For each wave number $\wavenum$, we construct the stability matrix and study its spectrum. To first order:

\def \nabsz{\ensuremath{\nabla_{sz}} \xspace}
\def \nabphi{\ensuremath{\nabla_{\phi}} \xspace}

\myequationn{
\dot{\temp}\rho = \bracket{-\airflow \mycdot \nabsz + \nabsz \mycdot \cond \nabsz -\cond   \paren{\frac{\wavenum}{s}}^{2} + \dens \metab_{\temp}} \MyDelt \temp + \\
\bracket{\metab \MyDelt \dens + \nabsz \cond_{\dens} \nabsz \temp} \MyDelt \dens -  \MyDelt \airflow  \mycdot \nabsz \temp\\
\phantom{a} \\
\dot{\dens} = \nabsz \mycdot \paren{\dens \nabsz \MyDelt \beepress} - \paren{\frac{\wavenum}{s}}^{2}  \dens \MyDelt \beepress\\
\MyDelt \beepress =   \frac{d \beepress}{d \dens} \MyDelt \dens +  \frac{d \beepress}{d\temp} \MyDelt \temp,
}

where $\nabsz$ is the gradient in the $s$ and $z$ directions, $\metab_{\temp} = \frac{d \metab(\temp)}{d\temp}$, $\cond_{\dens} = \frac{d \cond(\dens)}{d\dens}$. Regions where $\dens = \maxdens$ are locked at $\maxdens$ and not allowed to vary in bee packing fraction. 

\subsubsection{Solving for $\MyDelt \airflow$, $\MyDelt \press$}
The above equations depend on $\MyDelt \airflow$, which is determined by:
\myequationn{
\MyDelt \airflow = \darcy \bracket{\MyDelt \temp \hat{z} - \darcy \nabla \paren{\Delta \press}} + \Delta \dens \bracket{\paren{\temp - \tamb}\hat{z} - \nabla \press}.
}
$\MyDelt \airflow$ must have the form:

\myequationn{
\MyDelt \airflow = \bracket{ \MyDelt \airflow_{sz} -  \darcy \nabphi \MyDelt \press \hat{\phi} } = \bracket{ \MyDelt \airflow_{sz} -  \darcy \frac{i \wavenum}{s} \MyDelt \press \hat{\phi} },
}

where $\MyDelt \airflow_{sz}$ is the $sz$ component of $\MyDelt \airflow$. We solve for pressure $\MyDelt \press$ using the condition $\nabla \mycdot \MyDelt \airflow = 0$:

\myequationn{
\nabla \mycdot \MyDelt \airflow =  \nabsz \mycdot \airflow_{sz}+\nabphi \mycdot \airflow_{\phi} = \nabsz \mycdot \airflow_{sz} - \darcy \MyDelt \press \paren{\frac{i \wavenum}{s}}^{2} = 0 \Rightarrow \MyDelt \press  \darcy \paren{\frac{\wavenum}{s}}^{2} = {-\nabsz \mycdot \airflow_{sz}}{}.
}
At $\wavenum = 0$, $\MyDelt \airflow_{\phi} = 0$, this condition simply becomes $\nabsz \mycdot \airflow_{sz} = 0$.\\

Therefore, at a given $\MyDelt \temp, \MyDelt \dens$, we can solve $\MyDelt \airflow$, $\MyDelt \press$ from this set of linear equations.
\subsubsection{Numerical computation of linear response matrix}

To solve for stability, we discretize the system in the $s, z$ directions in the same way that we did when solving for the equilibrium state. Because we are only solving for linear stability and the system starts off uniform in the $\phi$ direction, we don't need to discretize in the $\phi$ direction. For perturbations at a certain wavenumber $\wavenum$, the temperature derivative is, to first order:

\myequationn{\frac{d\temp_{\ij}}{dt}  \dens_{\ij}   \myinvphase = \underbrace{\metab\paren{\bracket{\temp + \mydelt \temp}_{\ij} }\paren{\dens + \mydelt \dens}_{\ij} }_{\text{Metabolism}} - \\
 \underbrace{\mydelt \temp_{\ij} \paren{\frac{\wavenum}{s_{\ij}} }^{2}\cond(\dens_{\ij})}_{\text{Conduction in $\phi$ direction}}
%Conductivity
-\underbrace{\frac{1}{V_{\ij}}  \sum_{\expt{\ijp}} \frac{A_{\ijij}}{\cellw}   \harmeanb{\cond \paren{{\dens_{\ij} + \mydelt \dens_{\ij}}}}
{ \cond \paren{\dens_{\ijp} + \mydelt \dens_{\ijp}}}  \bracket{ \paren{\temp + \mydelt \temp}_{\ijp} -  \paren{\temp + \mydelt \temp}_{\ij}   }}_{\text{Conduction in $s, z$ directions}}  - \\
%Convection
+ \underbrace{\frac{1}{V_{\ij}}   \sum_{\expt{\ijp}}{ \paren{\discflow_{\ijij} + \mydelt \discflow_{\ijij}}   \bracket{-\theta  \paren{ \discflow_{\ijij} + \mydelt \discflow_{\ijij}}  \paren{\temp_{\ij}  + \mydelt \temp_{\ij}}+ \theta  \paren{- \bracket{\discflow_{\ijij} + \mydelt \discflow_{\ijij}}} \paren{\temp_{\ijp}  + \mydelt \temp_{\ijp}}}}}_{\text{Air flow in $s,z$ directions}} \\
+ \underbrace{\frac{1}{V_{\ij}}   \sum_{\expt{\ijp}}{\mydelt \discflow_{\ijij}}  \temp_{\ij}}_{\text{Air flow in $\phi$ direction}}.
} 
Note the slight modification of the upwinding terms, where we have also included a $\phi$ component to represent the influx or outflux of air in the $\phi$ direction. \\

The density derivative is, to first order:
\myequationn{ \frac{d \dens_{\ij}}{dt} \myinvphase  =  \beeflow_{0}  \bracket{ 
\underbrace{\paren{\frac{\wavenum}{s_{\ij}} }^{2}\dens_{\ij}   \paren{\beepress + \mydelt \beepress}_{\ij}}_{\text{Movement in $\phi$ direction}}
%Conductivity
-\underbrace{\frac{1}{V_{\ij}}  \sum_{\expt{\ijp}} \frac{A_{\ijij}}{\cellw}   \bracket{\frac{\dens_{\ij} + \dens_{\ijp}}{2}}  \bracket{\paren{\beepress + \mydelt \beepress}_{\ij} -\paren{\beepress + \mydelt \beepress}_{\ijp} }}_{\text{Movement in $s, z$ directions}}
}.
%Convection
}

The airflow is solved using the set of linear equations:

\myequationn{
\discflow_{\ijij}+ \mydelt \discflow_{\ijij}= \bracket{ \harmean{\darcy(\dens_{\ij}+ \mydelt \dens_{\ij})  }{\darcy(\dens_{\ijp} + \mydelt \dens_{\ijp})}  \frac{A_{\ijij}}{\cellw} }  \\
 \bracket{ \bracket{\paren{\press + \mydelt \press}_{\ij} - \paren{\press + \mydelt \press}_{\ijp}}   + \paren{z_{\ijp}-z_{\ij}} \paren{\frac{\temp_{\ij} + \mydelt \temp_{\ij} +  \temp_{\ijp} + \mydelt \temp_{\ijp}}{2}  }}.
}
$\mydelt \press$ is set such that the divergence in the $\phi$ direction negates the divergence in the $s, z$ directions,  
\myequationn{
 { \sum_{\expt{\ijp}}{\mydelt \discflow_{\ijij}}} =
  \mydelt \press_{\ij} \darcy(\dens_{\ij}) w^{2} \paren{\frac{\wavenum^{2}}{s_{\ij}}} }
 for all cells $(i, j)$.

%
%
%
% 
%
%A metabolic equivalent(MET) is $3.5 \ml \text{O}_{2}/ \text{kg min} = 4.184 \kJ/(\kg \mycdot \text{hour}) = 1.16 \watt/\kg = 1.16 \mycdot 10^{-3} \mycdot \watt/\ml$\\
%
%Heinrich uses units of $ml \text{O}_{2}/ (\gram \hour)$ which is 4.76 MET  \cite{Heinrich:1981ws}. His ranges from 1 MET to 20 MET, but it hovers more around 4-5. 
%Southwick has a density of roughly 0.5, radius of 9.5(Fairly 2DIsh). Their 608 \gram cluster is 6.46 $\ml/\minute$ which is $10.65 \ml/(\kg \mycdot \minute)$ which is 3 MET, which is $3.5 \mycdot 10^{-3} \mycdot \watt/\ml$.
%
%
%So for Southwick \cite{Southwick:1985eb}, the temperature is $34 - 23 \mycdot \paren(r/9.5)^{2}$ Given that it's in two dimensions, the laplacian is $1.02 \degc/\cm^{2}$. We know that $\metab \dens  = \cond \nabla^{2}\temp$. 
%$\cond = \frac{1.75 10^{-3} \mycdot \watt/\ml}{1.02 \degc/\cm^{2}}  = 0.0017 \watt/(\cm \degc)$
%%(1.75 * 10^(-3) watt/(cm^3))/(1.02 kelvin/(cm^2)) in watt/(cm * kelvin)
%
%
%For reference, dry air has a conductivity of $2.6 \mycdot  10^{-4}\watt/(\cm \degc)$. This means that fully packed, conductivity approaches dry air. 
%%http://www.engineeringtoolbox.com/dry-air-properties-d_973.html
%%
%

  \end{document}